# Shear viscosity and wall slip behavior of dense suspensions of polydisperse particles*

Jing He[1,2] , Stephanie Lee[2] and Dilhan M. Kalyon[1-3]

Highly Filled Materials Institute[1]

Chemical Engineering and Materials Science[2]

Biomedical Engineering[3]

Stevens Institute of Technology

Castle Point St. Hoboken, NJ 07030, USA

**Abstract**

Shear viscosity and wall slip of dense suspensions of a silicone polymer incorporated with polydisperse particles were investigated. Three types of particles with low aspect ratios were used to achieve a relatively high maximum packing fraction, $\phi_m$=0.86. Such a high $\phi_m$ allowed the preparation of suspensions with a wide range of solid volume fractions, $\phi$, i.e., $0.62 \leq \phi \leq 0.82$. The wall slip velocities of the suspensions in steady torsional and capillary flows were characterized and determined to be fully consistent with the mechanism of apparent slip layer formation at the wall. Upon wall slip corrections it was found that at shear stresses which are significantly above the yield stress the relative shear viscosity of the suspensions obeys well the Krieger-Dougherty relationship that links the relative shear viscosity behavior of dense suspensions solely to $\phi/\phi_m$. However, at lower shear stresses that are in the vicinity of the yield stresses the relative shear viscosity becomes functions of both $\phi/\phi_m$ and the shear stress. It is clearly demonstrated that without wall slip analysis the accurate characterization of the relative shear viscosity of dense suspensions is not possible.

# I.    INTRODUCTION

Concentrated suspensions of densely packed rigid particles in Newtonian or non-Newtonian fluids, i.e., "dense" suspensions, are processed widely in myriad industries including food, personal care, energetics, ceramics, pharmaceutical, magnetics, and composites industries. The characterization of the rheological material functions of dense suspensions is a challenge even when particles with low aspect ratios are used. The challenge stems from the difficulty of mixing the ingredients of suspensions under well-defined and reproducible conditions, as well as the viscoplasticity and wall slip behavior of dense suspensions.

The thermo-mechanical history that the suspension is exposed to during processing dictates the homogeneity of the spatial distribution of the particles and consequently the resulting rheological and wall slip behavior [1-6]. Ultimate properties including electrical properties [4, 7, 8], fire retardance [9], crosslink density and swelling [10], energetic properties [11] and the development of crystalline morphologies following shearing [12] are dependent on  the dynamics of the mixing processes.

A wide range of experimental data was collected on the flow behavior of dense suspensions [13-55]. Furthermore, various mathematical models and simulations of the flow and deformation behavior of dilute and concentrated suspensions were carried out to predict their shear viscosity [56-70]. However, in spite of such intense focus a number of fundamental questions still remain, including how the shear viscosity relates to the volume fraction, $\phi$, and the maximum packing fraction, $\phi_m$, of their particles when the suspensions are viscoplastic and exhibit wall slip.

Here we report on the viscometric flows of a silicone polymer (poly(dimethyl siloxane)), PDMS, incorporated with non-colloidal particles by especially focusing on the roles played by wall slip and yield stress formation. Polydisperse particles with relatively low aspect ratios were used to furnish a relatively high maximum packing fraction of the solid phase, $\phi_m = 0.86$. Such a relatively high maximum packing fraction enabled the preparation of dense suspensions with a wide range of concentrations, i.e., $0.62 \leq \phi \leq 0.82$. The suspensions were prepared under industrially-relevant batch processing conditions that generated similar mixing indices (as will be defined later) of suspensions with different $\phi$. Data from steady torsional and capillary flows were collected as functions of the surface to volume ratios of the rheometers so that the wall slip behaviors of the suspensions could be characterized and used to correct the flow curves of the suspensions. The relative shear viscosity values of the suspensions obtained upon wall slip velocity corrections were compared with the predictions of various theories.

## II.     EXPERIMENTAL

### 2.1     Rheological characterization

The rheological characterization of the PDMS and the suspension samples were carried out at the ambient temperature (22 ±0.2 °C) using small-amplitude oscillatory shear, steady torsional and capillary flows.

**Small-amplitude oscillatory shear:** The linear viscoelastic material functions of the PDMS binder fluid and the suspensions of PDMS were documented using an Advanced Rheometric Expansion System (ARES) rheometer available from TA Instruments of New Castle, DE. This rotational rheometer has a force rebalance transducer 0.2K-FRTN1 and was used with stainless steel parallel disks with 25-50 mm diameters. The torque accuracy of the transducer is ±0.02 g-cm. The actuator of the ARES is a DC servo motor with a shaft supported by an air bearing with an angular displacement range of $5 \times 10^{-6}$ to 0.5 rad, and angular frequency range of $1 \times 10^{-5}$ to 100 rad/s. The angular velocity range is $1 \times 10^{-6}$ to 200 rad/s. The sample loading procedure was kept the same for all the experiments. Consistent with our earlier investigations the suspension samples were not presheared. Preshearing itself can significantly change the structure and hence the resulting rheological behavior of concentrated suspensions [8].

During oscillatory shearing the shear strain, $\gamma$, varies sinusoidally with time, t, at a frequency of $\omega$, i.e., $\gamma(t) = \gamma^0 \sin(\omega t)$ where $\gamma^0$ is the strain amplitude. The shear stress, $\tau(t)$ response of the fluid to the imposed oscillatory deformation consists of two contributions associated with the energy stored as elastic energy and energy dissipated as heat, i.e., $\tau(t) = -G'(\omega)\gamma^0 \sin(\omega t) - G''(\omega)\gamma^0 \cos(\omega t)$ The storage modulus, $G'(\omega)$, and the loss modulus, $G'(\omega)$, also define the magnitude of complex viscosity, $|\eta*| = \sqrt{\left(\left(G'/\omega\right)^2 + \left(G''/\omega\right)^2\right)}$, and $\tan\delta = G''/G'$. In the linear viscoelastic region all dynamic properties are independent of the strain amplitude, $\gamma^0$.

**Steady torsional flow:** The parallel disk torsional flow experiments were again carried out using the ARES rheometer. Parallel disks with diameters of 25-50 mm were employed. The gap heights were varied in the 0.5 to 2 mm range for the determination of wall slip velocities as functions of shear stress [29, 33, 71].

**Capillary flow:** The capillary flow experiments were carried out using an Instron capillary rheometer in conjunction with an Instron Floor Tester and multiple capillaries with different length over diameter ratios and diameters [29, 33]. The length/diameter, L/D, ratios of the capillaries were varied between 20 and 60 at constant diameter for the correction of the wall shear stress under fully-developed flow conditions (Bagley correction). The diameters of the capillaries were varied in the 0.8 and 2.5 mm range at constant L/D ratio for enabling the determination of wall slip velocities and the correction of the wall shear rates for slip [15, 29, 72].

## 2.2 Materials

**The binder:** A poly(dimethyl siloxane) (PDMS) DMS-T35 available from Gelest, Inc. (density of 973 kg/m$^3$ at 25 °C) was used. We have chosen to use PDMS as the binder since silicone polymers have attracted special attention to their rheological behavior [73, 74]. The molecular weight of the PDMS was 48 kDa. The use of a PDMS with such a relatively low molecular weight was to avoid the wall slip of the binder itself. To verify that the PDMS itself does not exhibit wall slip, systematic experiments on pure PDMS at various surface to volume ratios of the rheometers were carried out. The PDMS of our investigation did not exhibit wall slip, consistent with the general understanding that PDMS resins generally exhibit wall slip when molecular weights >500 kDa (and typically when the wall shear stress exceeds 70 KPa [73, 75, 76]). The wall slip of the binder itself would have seriously complicated our analysis for the characterization of the shear viscosity of dense suspensions and thus was best avoided.

In order to ensure that the PDMS would be stable during characterization of its dynamic properties, the oscillatory shear data were first collected as a function of time at constant $\gamma^0$ and $\omega$. These time scans suggested that the PDMS was stable. The strain amplitude, $\gamma^0$, range associated with the linear viscoelastic behavior of PDMS was determined to be typically < 100% (Supplemental figures, Fig. S1). The dynamic properties of the PDMS as a function of $\omega$ are shown in Fig. 1. At the low deformation rates associated with linear viscoelasticity the behavior of the PDMS approaches the behavior of a Newtonian fluid, i.e., $G'' >> G'$ and $|\eta^*| \sim 4.5$ Pa-s.

Fig.2 shows the shear viscosity values of the PDMS collected over the shear stress range of 25-38000 Pa. The shear viscosity data of the PDMS binder at the lower shear stresses (25-2300 Pa) were collected with steady torsional flow, while the shear viscosity values at the higher shear stress end (3300-40000 Pa) were collected using capillary flow. The shear viscosity of the PDMS stays independent of the shear stress up to 1000 Pa, exhibiting a zero-shear viscosity, $\eta_0$, of 4.9 Pa-s, consistent with $|\eta^*|$ values. The shear viscosity of the PDMS starts to decrease significantly at shear stresses>3000 Pa following a power law fluid behavior (Ostwald-de Waale) with a power law index of 0.8 and consistency index of 17.9 Pa-s$^{0.8}$ (Fig. 2). During capillary flow experiments a thermal imaging camera (PM190, FLIR) was used to determine the temperature of the extrudate at its surface as it emerges from the capillary. The measured surface temperatures of PDMS extrudates were used to assess the importance of the viscous energy dissipation effects during capillary flow. Viscous energy dissipation was found to be negligible in spite of the relatively large shear stresses applied at the wall.

**Particles with a polydisperse size distribution**: A set of rigid particles with a polydisperse size distribution was employed. The largest particles consisted of smooth glass spheres (A2429, Potters Industries LLC., Malvern, PA, USA) with a mean diameter of 90 µm and specific gravity of 2.5 at ambient temperature. The medium size particles also consisted of smooth spherical borosilicate particles, with a mean diameter of 15 µm and a specific gravity of 1.09 (110P8, Potters Industries LLC., Malvern, PA, USA). The third set of particles, "small particles", exhibited low aspect ratios with some surface roughness, i.e., Dechlorane Plus 35 (Occidental Chemical Corp., Dallas, TX) with a mean characteristic length of 3.2 µm and a specific gravity of 1.8. The use of particles with rough surfaces was intentional and aimed at generating measurable yield stress values.

The volume fraction ratios of large, medium and small size particles were kept constant for all $\phi$ at 9:3:1 to generate a relatively high maximum packing fraction, $\phi_{\mathrm{m}}$. The method of Ouchiyama and Tanaka [77] was applied for the calculation of the maximum packing fraction as $\phi_{\mathrm{m}} = 0.86$ (Appendix).

The typical particle Reynolds number, $Re$, Peclet number, $Pe$, and Shear-Attraction number, $N_{SA}$, [78] were determined to be:

$$Re(\dot{\gamma}) = \frac{\rho_0 a^2 \dot{\gamma}}{\eta_0} \ll 1, \qquad Pe(\dot{\gamma}) = \frac{6\pi \eta_0 a^3 \dot{\gamma}}{\kappa T} \gg 1, \quad N_{SA}(\dot{\gamma}) = \frac{\eta_0 \dot{\gamma} a^3}{A} \gg 1$$

where $\rho_0$ is the binder density, $\dot{\gamma}$ is the true shear rate, $\eta_0$ is the zero shear viscosity of the binder, $a$ is the average radius of spherical particles, $\kappa$ is the Boltzmann constant, $T$ is the absolute temperature and $A$ is the composite Hamaker constant representing the interaction between the filler particles and the binder ($A \approx 0.2\kappa T$ for PDMS and glass particles [79]).

The true shear rates observed in parallel plate and capillary rheometry are in the range of $1 \times 10^{-3} \leq \dot{\gamma} \leq 1 \times 10^{3}$, i.e., a range which gives rise to a vanishingly small Reynolds number, i.e., $Re \leq 5 \times 10^{-7}$, indicating that all the flows were in the creeping flow regime. $Pe$ and $N_{SA}$ (using $A$ for PDMS and glass) were determined to be in the range of $1 \times 10^{3} \leq Pe \leq 4.9 \times 10^{5}$ and $2.6 \times 10^{2} \leq N_{SA} \leq 1.3 \times 10^{5}$ for the low shear rate region of steady torsional flow and $2 \times 10^{7} \leq Pe \leq 6.8 \times 10^{8}$ and $25.2 \times 10^{6} \leq N_{SA} \leq 1.8 \times 10^{8}$ for the high shear rates achieved with capillary rheometry, indicating that the viscous forces dominated over the colloidal forces as well as the attractive forces over the entire shear stress range of our experiments.

## 2.3    Wettabilities and mixing of the binder with the trimodal particles

**Wettability:** The wettabilities of the particles by the PDMS were characterized using a goniometer in conjunction with the sessile drop method (Appendix). The typical equilibrium contact angles of PDMS on films of glass and Dechlorane Plus were determined as 15° and 10°, respectively. Such relatively low contact angles suggest that the particles would be readily wet by the binder, to facilitate the facile dispersion of the particles within PDMS.

**Mixing:** An intensive batch mixer/torque rheometer, with a mixing volume of 300 mL, (EU-5V manufactured by Haake Buchler Instruments, Inc., Saddle Brooke, NJ, USA) was used to mix the particles with the binder. This mixer has two intermeshing counter-rotating rotors and can impart relatively high stress magnitudes (the design of the intensive mixer is that of the Banbury mixer that is widely used for the compounding of elastomers in the rubber industry). Mixing was carried out at ambient temperature. The rotational speed of the rotors was kept at 32 rpm and the degree of fill of the mixer (volume occupied by the suspension over the total available volume in the mixer) was 0.8. For each suspension the mixing time was systematically varied (5 to 30 minutes). Sets of five specimens each were collected at regular time intervals of 5-10 minutes for the determination of mixing indices. The suspensions were mixed under conditions which gave rise to relatively high mixing index values (as described next).

## 2.4 Mixing index characterization

It is challenging to predict optimum processing conditions for various batch or continuous processing equipment [80-90]. Alternatively, mixing geometries and conditions can be varied, while mixing indices are characterized, to determine optimum mixing conditions. Mixing indices allow the documentation of the statistics of the spatial concentration variation of the ingredients of dense suspensions so that the suspensions can be reproducibly prepared [91]. It is desirable to achieve similar mixing indices for suspensions with various $\phi$.

Mixing indices can be based on the measurement of concentration variations within the mixture, via various methods including wide-angle x-ray diffraction and/or thermo-gravimetric analysis [92-94]. Mixing indices can provide a quantitative understanding of the effects of mixing dynamics and document the effects of the geometries and operating conditions that are used during batch and continuous mixing of concentrated suspensions on the homogeneity of ingredient distributions [12, 95].

The following analysis was used to characterize the statistics of the concentration distributions of the filler particles to provide measures of the "degree/goodness of mixedness" or "mixing indices". If N measurements of the concentration, $c_i$, of one of the ingredients of the suspension formulation are made, then the mean, $c$ and the variance, $s^2$, of the concentration distribution of this particular ingredient are:

$$\overline{c} = \frac{1}{N}\sum_{i=1}^{N}c_i \quad \text{and} \quad s^2 = \frac{1}{(N-1)}\sum_{i=1}^{N}\left(c_i - \overline{c}\right)^2 \qquad \text{Eq. (1)}$$

A small variance value would suggest that the mixture approaches the behavior of a homogeneous system, where most of the samples yield concentration values, $c_i$, that approach the mean concentration, $\overline{c}$. On the other hand, the poorest mixing state would pertain to the components of a mixture being completely segregated from each other. The value of the maximum variance $s_0^2$ for a segregated system can be defined by assuming that the samples are taken from either one component or the other without crossing a boundary (Eq. (2)). The degree of mixedness, i.e., mixing index, *MI*, can be defined as the ratio of the standard deviation of the distribution over the standard deviation of the completely segregated sample as:

$$s_0^2 = \overline{c}\left(1 - \overline{c}\right) \quad \text{and} \quad MI = 1 - \frac{s}{s_0} \qquad \text{Eq. (2)}$$

Thus, the value of the mixing index, *MI*, would approach one for a completely random distribution of the ingredients and to zero for the segregated state.

For the mixing index determination the suspension samples were subjected to thermo gravimetric analysis, TGA, using a TA Instruments Q50. Five specimens of around 20 mg each were collected from different locations inside the mixer. Each specimen was heated at a rate of 20 °C/min from ambient to 550 °C and held at this temperature for 20 min. Weight changes were recorded as a function of time and temperature. The TGA of pure ingredients revealed that the PDMS and Dechlorane Plus completely decompose below 415 °C. The borosilicate glass particles were determined to be stable up to the 550 °C. Consequently, the total concentration variation of the glass particles was used for the determination of the mixing indices that are reported in Fig. 3.

The inset of Fig. 3 shows typical mixing index, *MI*, values as a function of mixing time for $\phi$=0.76. The mixing index reaches a constant value which approaches 1.0 above a critical time of mixing, suggesting that a homogeneous spatial distribution of the glass particles occurs (at the scale of examination of 20 mg sample size). Critical times to achieve similarly high mixing index values could be obtained for the suspensions with other $\phi$ (except for $\phi$=0.82 as shown in Fig. 3). These mixing index, *MI*, values > 0.995, can be considered to reflect well-distributed glass particles. For $\phi$=0.82 the *MI* could only be increased up to 0.990 regardless of the mixing conditions.

## III.   BACKGROUND ON WALL SLIP ANALYSIS

**Apparent slip mechanism:**

During the flow of a suspension of rigid particles, the particles cannot physically occupy the space adjacent to a wall as efficiently as they can away from the wall. This leads to the formation of a generally relatively thin, but always present, layer of fluid adjacent to the wall, i.e., the "apparent slip layer" or the "Vand layer" [16]. It can be assumed that the apparent slip layer consists solely of the binder of the suspension and itself adheres to the wall, that its thickness, $\delta$, is sufficiently small so that the separation of the binder from the bulk suspension to form the slip layer does not affect the shear viscosity of the suspension, and $\delta$ is not affected by channel gap or the volumetric flow rate (except for the plug flow region) [72, 96].

**Slip layer thickness from extruded strands**

The thicknesses of the apparent slip layer, $\delta$, of extrudates emerging from capillary flow was characterized employing cryo-scanning electron microscopy (cryo-SEM) and energy dispersive X-ray spectroscopy (EDX). The details of the apparatus and the procedures used are given in Appendix.

**Steady torsional flow:**

The apparent shear rate, $\dot{\gamma}_a$, (not corrected for slip effects) is a linear function of the radial distance from the center of the disk, r, as given by:

$$\dot{\gamma}_a = \frac{\Omega r}{H} \qquad \text{Eq.(3)}$$

where $H$ is the gap height, and $\Omega$ is the angular velocity of the bottom disk. The shear stress at the edge of the disk, $\tau_R$, can be determined from:

$$\tau_R = \frac{\Im}{2\pi R^3}\left(3 + \frac{d\ln\Im}{d\ln\dot{\gamma}_{aR}}\right) \qquad \text{Eq.(4)}$$

where $\Im$ is the torque generated by rotating the bottom disk and $\dot{\gamma}_{aR}$ is the apparent shear rate at the edge of the disk. The true shear rate at the edge, $\dot{\gamma}_R\left(\tau_R\right)$ is defined as [71]:

$$\dot{\gamma}_{aR} = \dot{\gamma}_R\left(\tau_R\right) + \frac{2U_s\left(\tau_R\right)}{H} \qquad\qquad \text{Eq. (5)}$$

Eq. 5 suggests that if plots of apparent shear rate versus reciprocal gap are drawn at constant shear stress at the edge, then straight lines would be obtained. The extrapolated intercepts would be equal to the true shear rate at the edge, and the slopes would be equal to $2U_s\left(\tau_R\right)$.

Under apparent slip conditions the relationship between the slip velocity, $U_s$, and the shear stress, $\tau_R$, becomes the following in steady torsional flow for any binder, the shear viscosity of which can be represented by a power law equation, i.e., $\tau_R = m_b\,\dot{\gamma}_R^{\,n_b}$:

$$U_s\left(\tau_R\right) = \pm\frac{\delta}{m_b^{\,1/n_b}}\tau_R^{\,1/n_b} = \pm\beta\tau_R^{\,s_b} \qquad\qquad \text{Eq. (6)}$$

here, $\pm$ is necessary to accommodate the changing sign of the slip velocity (velocity of the fluid at the wall minus the wall velocity) at the stationary and moving disks, $\delta$ is the slip layer thickness, $\beta$ is the slip coefficient (defined as Navier's slip coefficient for a linear relationship between $U_s\left(\tau_R\right)$ and $\tau_R$) and $m_b, n_b$ are the consistency index and the power law index parameters of the "power-law" type shear viscosity of the binder and $s_b = 1/n_b$.

During the steady torsional flow experiments, a straight marker line was placed to cover the edges of the fixtures and the free surface of the specimen. This marker line technique can be used for the characterization of the wall slip behavior of concentrated suspensions in steady torsional flow [33, 97]. Here, it was used only to validate the wall slip velocities which were determined from systematic experiments carried out by varying the surface to volume ratio during steady torsional flow.

**Capillary flow:**

In capillary flow the shear stress at the wall, $\tau_w$, corresponding to fully-developed flow can be determined by correcting for the pressure losses associated with end effects [98], i.e., $\tau_w = \dfrac{\Delta P D}{4(L + ND)}$, where $\Delta P$ is the total pressure drop over the capillary, $L$ is the length and $D$ is the diameter of the capillary, and $N$ is the equivalent length associated with the end correction (entrance and exit effects). The equivalent length can be determined by the extrapolation of the pressure drop versus the length over diameter ratio curve to intersect with the length over diameter axis. The apparent shear rate at the wall, $\dot{\gamma}_{aw} = 8\overline{V}/D$, where $\overline{V}$ is the mean fluid velocity. To determine the wall slip velocity values, capillaries with the same length over diameter ratios but with different diameters were used. The analysis proposed by Mooney for fully developed, incompressible, isothermal, and laminar flow in circular tubes with a slip velocity of $u_s$ at the wall yields [15]:

$$\frac{8\overline{V}}{D} = \frac{4}{\tau_w^3} \int_0^{\tau_w} \tau^2 \dot{\gamma} \, d\tau + \frac{8U_s}{D} \qquad \text{Eq.(7)}$$

where $\tau$ is the shear stress and $\dot{\gamma}$ is the true shear rate. Differentiating the last equation with respect to $1/D$ at constant shear stress at the wall, $\tau_w$, one obtains:

$$\frac{\partial\left(8\bar{V}/D\right)}{\partial\left(1/D\right)}\bigg|_{\tau_w} = 8U_s \qquad \text{Eq. (8)}$$

Thus, the plot of the apparent shear rate at the wall, $\left(8\bar{V}/D\right)$, versus $1/D$ at constant $\tau_w$ should give a straight line with a slope of $8U_s$. The slip corrected wall shear rate, $\dot{\gamma}_w$, of the suspension is given as [30, 72]:

$$\dot{\gamma}_w = \frac{\left(Q-Q_s\right)}{\pi R^3}\left[3+\frac{\text{d}\ln\left(Q-Q_s\right)}{\text{d}\ln\tau_w}\right] \qquad \text{Eq.(9)}$$

where, $R$ is the capillary radius, $Q$ is the volumetric flow rate at wall shear stress of $\tau_w$ and $Q_s$ is the volumetric flow rate due to slip, i.e., $Q_s = \left(\pi/4\right)D^2U_s$. The slip velocity for capillary flow, $U_s$, under apparent slip flow conditions is given as [72] :

$$U_s = \beta\tau_W^{s_b}, \text{ with } \beta = \frac{R}{m_b^{s_b}\left(s_b+1\right)}\left[1-\left(1-\frac{\delta}{R}\right)^{s_b+1}\right] \qquad \text{Eq.( 10a)}$$

where $\tau_W$ is the wall shear stress, $R$ is the capillary die radius, $\beta$ is the Navier's slip coefficient, $m_b, n_b$ are the parameters of the shear viscosity of the binder, $s_b = 1/n_b$, and $\delta$ is the slip layer thickness. Considering that the reciprocal power-law index, $s_b$, of the binder is positive, and assuming an integer value-the use of binomial theorem provides $\left(1-\dfrac{\delta}{R}\right)^{s_b+1} \cong 1 - \dfrac{\delta}{R}\left(s_b+1\right)$ thus, generating a slip coefficient, $\beta = \delta \big/ m_b^{s_b}$

so that:

$$U_s = \beta \tau_W^{s_b} = \left(\delta \big/ m_b^{s_b}\right)\tau_W^{s_b} \qquad\qquad \text{Eq. (10b)}$$

similar to that obtained from steady torsional flow.

## IV.  RESULTS AND DISCUSSION

For the suspensions the strain amplitudes at which linear viscoelastic behavior prevails are restricted to relatively small strain amplitudes, $\gamma^0$, i.e., $\gamma^0 \le 0.02\%$. With increasing $\phi$ the storage modulus, $G'$ and the magnitude of complex viscosity, $|\eta^*|$, values increase by about two orders of magnitude as packing of the particles becomes more difficult (Fig. 4). The slope of $G'$ with $\omega$ also decreases with increasing $\phi$.

As $\phi$ approaches $\phi_m$ the particles can form percolating networks leading to significant transitions in viscoelastic properties from being based on polymer chain entanglement dynamics to a frequency insensitive gel-like behavior. The gel-like behavior is induced by particle to particle interactions which  results in the development of particle to particle networks and yield stresses [19, 44, 91]. Upon the adoption of gel-like behavior concentrated suspensions no longer manifest the typical die swell ("extrudate swell"), first normal stress difference, and rod climbing behavior of their polymeric binders. As $\phi$ increases, the die swell and the first normal stress difference values decrease monotonically [44]. With increasing $\phi$ the die swell values are eventually reduced to those of Newtonian fluids (swell values between 0.85 to 1.25) and the first normal stress difference values can reach negative values (leading to reversal of the rod climbing effect, i.e., dipping instead of climbing [44]). The development of a yield stress leads to overall shear thinning behavior since at shear stresses that are below the yield stress of the suspension the suspension is solid-like (exhibiting infinite shear viscosity). On the other hand, when the shear stress becomes greater than the yield stress particle to particle networks are broken and the suspension becomes fluid-like (finite shear viscosity).

Figure 5 shows the typical shear stress at the edge, $\tau_R$, versus the apparent shear rate at the edge, $\dot{\gamma}_{aR}$, data obtained from steady torsional flow for $\phi$=0.76. At a given apparent shear rate the shear stress increases with increasing gap and thus with decreasing surface to volume ratio. This dependency of the flow curves on the surface to volume ratio points to the important role that wall slip plays. The inset in Fig. 5 shows the apparent shear rate versus the reciprocal gap data at constant $\tau_R$ for $\phi$=0.76. The data points fall reasonably well on the linear regression lines drawn through the points. The slip velocity, $U_s(\tau_R)$ values are then determined from the slopes (slope= $2U_s(\tau_R)$) and are used to correct the flow curve [29, 71]. The slip corrected shear rates at the edge, $\dot{\gamma}_R$, are lower than the apparent shear rates at the edge, $\dot{\gamma}_{aR}$, i.e., $\dot{\gamma}_R < \dot{\gamma}_{aR}$, reflecting the important role wall slip plays.

The ratios of the slip velocity over the plate velocity, $U_s(\tau_R)/V_R$, versus the shear stress at the edge, $\tau_R$, at different gaps, H, obtained for $\phi$=0.76 are shown in Fig. 6. Plug flow is indicated when the ratio of slip velocity over the plate velocity, $U_s(\tau_R)/V_R$, is equal to 0.5 [72]. The yield stress of the suspension is the shear stress at which the transition from plug flow to the deformation of the suspension occurs [72]. Accordingly, the yield stress of the suspension, $\tau_0$, at $\phi$=0.76 can be estimated as 85 Pa. At shear stresses that are greater than the yield stress of the suspension ($\tau_R > 85$ Pa), the suspension yields as indicated by $(U_s(\tau_R)/V_R) < 0.5$ and exhibits "fluid-like" behavior, i.e., the suspension continues to deform as long as a shear stress that is greater than the yield stress is applied.

The plug-flow of the suspension is also demonstrated (inset of Fig. 6) via the placement of the straight marker line on the free surface of the suspension sample under plug flow conditions. An initial, relatively small deformation of the suspension is followed by lack of deformation, plug flow, during the rest of the motion of the top disk, suggesting that the shear stress imposed < yield stress [33, 42]. From the straight line marker technique the shear stress values at which there is no steady deformation of the suspension are successively determined to define the yield stress of the suspension. Using this method the yield stress at $\phi$=0.76 was determined to be within 60-100 Pa. This range is consistent with the yield stress value of 85 Pa obtained at $\phi$=0.76 from the slip velocity analysis (Fig. 6). Following similar procedures the slip velocity analysis for the suspension with $\phi$=0.78 generated a yield stress of 185 Pa (Supplemental Fig. S3), which was consistent with the yield stress value range determined from the straight marker line method for $\phi$=0.78, i.e., 150-200 Pa. For lower $\phi$ the yield stress values were not measured but were estimated using the yield stress values obtained for $\phi$=0.76 and 0.78 and the following relationship between the yield stress, $\tau_0$, and $\phi/\phi_m$.

$$\tau_0 = \frac{\alpha_1\left(\dfrac{\phi}{\phi_m}\right)}{\left(1-\dfrac{\phi}{\phi_m}\right)^{\alpha_2}} = \frac{0.06\left(\dfrac{\phi}{\phi_m}\right)}{\left(1-\dfrac{\phi}{\phi_m}\right)^{3.4}} \qquad \text{Eq. (11)}$$

Eq. (11) satisfies $\tau_0 = 0$ for $\phi = 0$ and $\tau_0 = \infty$ for $\phi = \phi_m$. The constants, $\alpha_1$ and $\alpha_2$ which best fit Eq. 11 were $\alpha_1 = 0.06$ and $\alpha_2 = 3.4$.

The slip velocity versus shear stress behavior of the suspensions characterized using steady torsional flow for $0.62 \leq \phi \leq 0.78$ are shown in Fig. 7. There is a linear relationship between the wall slip velocity and the shear, i.e., $U_s\left(\tau_R\right) = \beta\tau_R$. As noted earlier a linear relationship between $U_s\left(\tau_R\right)$ and $\tau_R$ (the slip exponent, $s_b$=1) is indicative of a binder with a constant shear viscosity over the shear stress range of the steady torsional flow experiments. As shown in Fig. 2 the PDMS binder indeed exhibits a constant shear viscosity up to 3000 Pa. The slip coefficient $\beta$ decreases with increasing $\phi$, consistent with an apparent slip layer thickness, which decreases with increasing $\phi$ (to be discussed later).

Fig. 8 shows capillary flow results for $\phi = 0.76$. In this figure the relationships between the wall shear stress, $\tau_w$, (end effect-corrected, i.e., Bagley correction applied) versus the apparent wall shear rate, $\dot{\gamma}_{aw}$, are provided for three capillaries with different diameters (0.8, 1.5 and 2.5 mm) at constant capillary length over diameter, $L/D$, ratios of 20 and 60. The inset shows the application of the Mooney procedure from which the slopes of the apparent shear rate versus the reciprocal diameters of the capillary dies are obtained. The slopes are equal to $8U_s(\tau_w)$ [15, 29]. Upon the application of the wall slip correction (Eq. (9)) wall shear stress, $\tau_w$, versus the slip-corrected wall shear rate, $\dot{\gamma}_w$, are obtained (Fig. 8). The effect of the wall slip is to reduce the shear rate at the wall in comparison to the apparent shear rate at the wall, $\dot{\gamma}_{aw}$. Consistent with steady torsional flow, the capillary flow results again attest to the importance of the applications of wall slip velocity corrections for the determination of the true shear viscosity behavior.

Similar capillary flow procedures and determinations of the wall slip velocities were carried out for $0.62 \leq \phi \leq 0.78$. However, capillary flow data could not be collected for $\phi \geq 0.80$ at which significant flow instabilities were observed. The flow instabilities were associated with the cyclic formation and destruction of solid mats at the converging flow zone at the entrance to the capillary die. The behavior was typical of formation of particle mats, accompanied by the filtration of the binder which generates a cyclic normal force on the plunger driving the flow [34, 99] (Supplemental figures, Fig. S4).

Fig. 9 shows the slip velocity values at the wall, versus the wall shear stress data collected with capillary flow. As noted earlier the basic mechanism of apparent slip formation suggests $U_s = \beta(\phi)\tau_w^{s_b}$ for a binder, the shear viscosity of which follows a power law relationship in the shear stress range of the experiments. Fig. 9 indicates that for capillary flow the slip exponent, $s_b$=1.25. Considering again that for the apparent wall slip mechanism $s_b$ should be equal to the reciprocal of the power law index, $n_b$, of the binder, i.e., $s_b = 1/n_b$, the power law index of the binder, $n_b$=1/1.25=0.8 is expected for the higher wall shear stress values of the capillary flow. Fig. 2 shows that the PDMS indeed exhibits a power-law behavior with $n_b$ =0.8 for $\tau_w$ > 3000 Pa.

The slip coefficients, $\beta$, determined with steady torsional and capillary flows can be used to determine the apparent slip layer thicknesses using $\beta = \dfrac{\delta}{m_b^{1/n_b}}$ . The slip coefficient $\beta$ values are tabulated in Table I along with the apparent slip layer thicknesses, $\delta(\phi)$). The $\delta$ values decrease significantly from 2.4-2.6 μm at $\phi$=0.62 to 0.3-0.7 μm at $\phi$=0.78. Previous studies have suggested that a linear relationship exists between the slip layer thickness over the harmonic mean particle diameter ratio, $\delta/D_P$, versus the ratio of the volume fraction of the particles over their maximum packing fraction, $\phi/\phi_m$ [72]:

$$\frac{\delta}{D_p} = 1 - \frac{\phi}{\phi_m}$$

Eq. (12)

Fig. 10 shows the apparent slip layer thicknesses determined from earlier investigations along with the slip layer thicknesses obtained here. As Fig. 10 shows Eq. (12) is also valid for the dense suspensions of this study.

The apparent slip layer thickness, $\delta$, obtained from the wall slip velocity data could be compared with δ that was determined directly from the EDX analysis of the cross-sectional areas of the extruded stands from capillary flow. The inset of Fig. 10 shows the typical identification of the apparent slip layer using the EDX method for $\phi$=0.72 collected at a wall shear stress of 40 kPa. In this method the glass spheres were identified by C subtracted Si mapping, labeled as diagonal shades, while Dechlorane Plus (3.5 μm) was identified by Cl mapping labeled as horizontal shades. The EDX method suggests that there is a distribution of the apparent slip layer thickness around the perimeter of the extrudate (depicted in black) with the thicknesses typically in the range of 0.5 to 3.5 μm. This range is not too far removed from the slip layer thickness, $\delta$, of 1.3 μm obtained from the apparent wall slip velocity data from capillary flow for $\phi$=0.72 (Table I).

The wall slip velocity data were used to correct the shear viscosity values of the suspensions for $0.62 \leq \phi \leq 0.82$ and the resulting shear viscosity values are shown in Fig. 11 for steady torsional flow and in Fig. 12 for capillary flow. The dense suspensions exhibit solid like behavior (particle to particle network formation) at shear stresses less than the yield stress at each $\phi$, followed by pseudo-plastic (shear thinning) behavior at shear stresses that are greater than the yield stress (Fig. 11). The shear viscosity of the suspensions can be represented using the Herschel-Bulkley Equation,

$$\eta(\phi, \tau_R) = m(\phi)^{1/n} \frac{\tau_R}{(\tau_R - \tau_0(\phi))^{1/n}} \text{ for } \tau_R > \tau_0 \text{ and } \eta(\phi, \tau_R) = \infty \text{ for } \tau_R \leq \tau_0 \text{ (for positive } \tau_R).$$

The parameters of the Herschel-Bulkley fluid for describing the flow curves at various $\phi$, obtained with steady torsional flow, are shown in Table II. The yield stress values were available from wall slip analysis (Fig. 6 and S3) and Eq. 11. Overall, as shown in Fig. 11 the fit is acceptable.

The relative shear viscosity (shear viscosity of the suspension over the shear viscosity of the binder, i.e., $\eta_r(\phi, \tau_R) = \eta(\phi, \tau_R)/\eta_b$) behavior of the suspensions at the relatively low shear stress range of the steady torsional flow experiments, i.e.,

$$\eta_r\left(\phi,\tau_R\right)=\eta\left(\phi,\tau_R\right)/\eta_b=\frac{m(\phi)^{1/n}}{\eta_b}\frac{\tau_R}{\left(\tau_R-\tau_0\left(\phi\right)\right)^{1/n}} \quad \text{for } \tau_R>\tau_0 \quad \text{(for positive } \tau_R\text{) are shown}$$

in Fig. 12. Viscoplastic behavior gives rise to a dependence of the relative shear viscosity of the suspension, $\eta_r$, not only on the $\phi/\phi_m$ ratio but also on the shear stress as shown in Fig. 12 for the relatively low shear stress ranges applied during steady torsional flow (up to 800 Pa). This finding might explain similar results reported by other investigations [57, 100-105] on the dependence of the relative viscosity on both the $\phi/\phi_m$ ratio and shear stress that were.

In contrast, the dense suspensions exhibit constant (shear stress independent) shear viscosity values at the higher wall shear stresses of capillary flow (Fig. 13). For capillary flow the wall shear stress values (4000-150,000 Pa) are significantly higher than the yield stress values of the suspensions (3.4-185 Pa for $0.62\leq\phi\leq0.78$) and thus the effects of the yield stress on shear viscosity eventually diminish to negligible levels. The relative viscosity, $\eta_r$, values (the Newtonian shear viscosity of the suspension obtained at the high shear stresses of capillary flow over the zero shear viscosity of the binder) are shown in Fig. 14. The relative shear viscosity, $\eta_r$, of the suspensions could be correlated solely with the $\phi/\phi_m$ ratio, i.e., $\eta_r$ was found to be independent of the shear stress. This was expected on the basis of the shear stress-independent shear viscosity values observed at the higher wall shear stresses of capillary flow (Fig.

13). It should again be noted that $\phi_m$ used in Fig. 14 is not an adjustable parameter and its value, i.e., $\phi_m = 0.86$, was obtained directly from the particle size distribution of the solid phase (Appendix).

A number of well-known expressions are available for representing the relative shear viscosity of concentrated suspensions. For example, Frankel and Acrivos have considered the simple shear flow of Newtonian suspensions consisting of monodisperse spheres incorporated into a Newtonian binder [58]. It was assumed that the particles move at the velocity of the fluid surrounding the particles, in close proximity and in relative motion with respect to each other. Viscous energy dissipation arising from the flow within the particle gaps separates the solid spheres. With this mechanism the relative shear viscosity of the suspension $\eta_r(\phi)$ depends on $\phi / \phi_m$ as [58]:

$$\eta_r(\phi) = \frac{\eta_s(\phi)}{\eta_b} = \frac{9}{8}\left[\frac{(\phi / \phi_m)^{1/3}}{1-(\phi / \phi_m)^{1/3}}\right] \qquad \text{Eq.(13)}$$

where $\eta_b$ and $\eta_s(\phi)$ are the Newtonian viscosities of the binder and the suspension, respectively. Another equation linking $\eta_r$ solely to $\phi / \phi_m$ is the Krieger and Dougherty (1959) expression:

$$\eta_r(\phi) = \frac{\eta_s(\phi)}{\eta_b} = \left(1-\frac{\phi}{\phi_m}\right)^{-k\phi_m} \qquad \text{Eq.(14)}$$

where $k$ is the intrinsic viscosity, i.e., a measure of the resistance to flow of a single particle. Factors that cause an increase in $k$ can result in a decrease of the maximum packing fraction, $\phi_m$, yet, the product remains in the range of 1.5 to 3. Various relevant approximations that are even applicable to anisometric solids (aspect ratios of up to 30) [25], to suspensions with polydisperse particles [31] or to aggregated systems [60] indicate that the product $k\phi_m$ remains constant and is equal to 2 [23, 105]. Equation (14) becomes:

$$\eta_r(\phi) = \frac{\eta_s(\phi)}{\eta_b} = \left(1 - \frac{\phi}{\phi_m}\right)^{-2} \qquad \text{Eq. (15)}$$

For our data shown in Fig. 14 the best agreement of the experimental values of $\eta_r(\phi)$ versus $\eta_r(\phi)$ obtained from Eq. 14 was obtained at a $k$ value of 2.5, i.e., $k\phi_m = 2.15$. The product $k\phi_m$ was therefore determined to be indeed close to 2. Thus, there is very good agreement of the $\eta_r(\phi/\phi_m)$ data with the predictions of the Krieger-Dougherty equation for the higher shear stresses of capillary flow. Overall, these findings are also consistent with the recent computational results of Pednekar *et al.* who have shown that for dense suspensions with particles exhibiting bimodal size distributions only a single parameter, $\phi_m$, is sufficient to represent their relative shear viscosity, $\eta_r$, versus $\phi/\phi_m$, behavior, i.e., consistent with the Krieger and Dougherty expression [70].

Can these findings be used to develop a methodology for *a priori* assessment of the significance of wall slip under different flow conditions? This is elucidated next using tubular flow. The total flow rate, $Q$, versus the wall shear stress, $\tau_w$, relationship for the apparent slip flow of a viscoplastic suspension with parameters $m$, $n$ and $\tau_0$ in a tube with radius, $R$, is [72]:

$$Q = Q_s + \frac{\pi R^3 \tau_w^{1/n}}{\left(\frac{1}{n}+1\right)\left(m(\phi)\right)^{1/n}} \left[ \left(1-\frac{\tau_0(\phi)}{\tau_w}\right)^{1/n+1} - \frac{2}{\left(\frac{1}{n}+3\right)}\left(1-\frac{\tau_0(\phi)}{\tau_w}\right)^{1/n+3} - \frac{2}{\left(\frac{1}{n}+2\right)}\frac{\tau_0(\phi)}{\tau_w}\left(1-\frac{\tau_0(\phi)}{\tau_w}\right)^{1/n+2} \right]$$

Eq. (16)

where $Q_s = \pi R^2 U_s = \pi R^2 \dfrac{\delta}{m_b^{1/n_b}} \tau_w^{1/n_b}$. Using the Binomial theorem for the approximation of the three terms in parentheses of Eq. 16 and $\delta = D_p\left(1-\dfrac{\phi}{\phi_m}\right)$ the ratio of the flow rate due to wall slip over the total flow rate, $Q_s/Q$, becomes:

$$\frac{Q_s}{Q}(\phi, \tau_w) = \frac{1}{1+\dfrac{\tau_w^{1/n-1/n_b} R \; m_b^{1/n_b}}{\left(m(\phi)\right)^{1/n} D_p \left(1-\dfrac{\phi}{\phi_m}\right)}\left[\dfrac{1}{(1/n+3)} - \dfrac{1/n}{(1/n+2)}\dfrac{\tau_0(\phi)}{\tau_w} + \dfrac{2}{(1/n+1)}\left(\dfrac{\tau_0(\phi)}{\tau_w}\right)^2\right]}$$

Eq. (17)

Plug flow is indicated for $Q_s/Q = 1$. For Bingham Plastic suspensions, i.e., n=1:

$$\frac{Q_s}{Q}\left(\phi,\,\tau_w\right)=\frac{1}{1+\left(\dfrac{R}{4D_p}\right)\dfrac{\tau_w^{\,1-1/n_b}\,m_b^{\,1/n_b}}{m(\phi)\left(1-\dfrac{\phi}{\phi_m}\right)}\left[1-\dfrac{4}{3}\left(\dfrac{\tau_0(\phi)}{\tau_w}\right)+4\left(\dfrac{\tau_0(\phi)}{\tau_w}\right)^2\right]}\qquad\text{Eq. (18)}$$

For the range of shear stresses at which the suspension behaves like a Newtonian fluid, i.e., $n=1$ and $\tau_0(\phi)=0$:

$$\frac{Q_s}{Q}\left(\phi,\,\tau_w\right)=\frac{1}{1+\left(\dfrac{R}{4D_p}\right)\dfrac{\tau_w^{\,1-1/n_b}\,m_b^{\,1/n_b}}{m(\phi)\left(1-\dfrac{\phi}{\phi_m}\right)}}\qquad\text{Eq. (19)}$$

and for Newtonian suspensions with Newtonian binders, with the relative shear viscosity of the suspension, $\eta_r(\phi)=\dfrac{\eta_s(\phi)}{\eta_b}=\left(1-\dfrac{\phi}{\phi_m}\right)^{-2}$:

$$\frac{Q_s}{Q}(\phi)=\frac{1}{1+\left(\dfrac{R}{4D_p}\right)\left(1-\dfrac{\phi}{\phi_m}\right)}\qquad\text{Eq. (20)}$$

Overall, the apparent slip mechanism of a suspension represented in Equations 16-20 suggests that as $\phi\to\phi_m$, $Q_s/Q\to1$, i.e., the flow approaches plug flow with increasing concentration of the rigid particles. Such plug flow behavior is indeed observed during the Poiseuille flow of Newtonian suspensions with increasing $\phi$, irrespective of the flow rate [106]. Apparent slip has a negligible effect for relatively small concentrations of rigid particles, i.e., as $\phi \rightarrow 0$ and $\tau_0 \rightarrow 0$, $Q_s/Q \rightarrow 0$ since $R \gg 4D_p$. As the radius, $R$, of the capillary increases, the contribution of the apparent wall slip decreases. Figure 15 shows some typical comparisons of the predictions of Equation (19) versus the experimental $Q_s/Q$ data for $\phi$=0.76. Similar results obtained for other $\phi$ are shown in the supplemental figure, Fig. S13. The general trend is that $Q_s/Q$ increases with increasing wall shear stress and decreasing capillary radius, $R$. Overall, the contribution that wall slip makes to the flow rate versus wall shear stress relationship is significant for all $\phi$. Fig. 15 suggests that it is indeed possible to rely on relatively simple expressions to describe apparent wall slip and shear viscosity of dense suspensions for the *a priori* assessment of the approximate contribution that wall slip is expected to make.

## V.    Conclusions

Particles with relatively low aspect ratios and polydisperse size distributions were used to provide a relatively high maximum packing fraction, $\phi_m$=0.86. This relatively high $\phi_m$ allowed the preparation of dense suspensions of a silicone polymer with a wide range of solid volume fractions, i.e., $0.62 \leq \phi \leq 0.82$. To keep consistent and reproducible mixing conditions the concentration variations were measured via thermo gravimetric analysis to allow the preparation of suspensions with similar mixing indices. The steady torsional and capillary flow data were collected as a function of the surface to volume ratio of the flow gaps to enable the determination of wall slip velocities.

The wall slip velocity values, $U_s(\tau_w)$, were found to be consistent with the mechanism of apparent slip layer formation at the wall, i.e., $U_s(\tau_w) = \beta \tau_w^{s_b} = \left(\delta / m_b^{s_b}\right) \tau_w^{s_b}$. The apparent slip layer thicknesses, $\delta$, depend on the $\phi / \phi_m$ ratio and the slip exponent, $s_b$, is equal to the reciprocal power law index of the binder, $1/n_b$.

Overall, the findings emphasize the important role that wall slip plays in the characterization of the relative shear viscosity behavior of dense suspensions. The wall slip effect has to be "unmasked" to arrive at the true shear viscosity behavior of the dense suspensions. For the determination of the parameters of constitutive behavior, again wall slip velocities need to be determined and used to arrive at the true parameters of viscoplastic constitutive equations of the suspensions. At shear stresses which are significantly above the yield stress the relative shear viscosity values of the suspensions obey well the Krieger-Dougherty relationship, linking the relative shear viscosity behavior of the suspensions solely to $\phi / \phi_m$. However, at lower shear stresses that are in the vicinity of the yield stresses the relative shear viscosity of the suspensions becomes functions of both $\phi / \phi_m$ and the shear stress.

**Appendix**

**Particle size distribution and maximum packing fraction:**

Fig. 16 shows the particle size distribution obtained through using scanning electron microscopy (Zeiss Auriga Small Dual-Beam FIB-SEM) along with computerized image analysis (ImageJ). The characteristic lengths of Dechlorane Plus are between 2 and 4 µm. The medium size glass spheres have a wider size distribution with diameters in between 6 and 26 µm. The diameters of the large glass spheres are distributed in between 42 and 116 µm with 85% of them between 80 and 110 µm.

We have used the theory of Ouchiyama and Tanaka [77] for the determination of the maximum packing fraction, $\phi_m$ [77, 107, 108]. This method involves the solution of the following equations:

$$\phi_m = \frac{\sum D_i^3 f_i}{\sum \left(D_i \sim D_a\right)^3 f_i + \dfrac{1}{\beta}\sum\left[\left(D_i + D_a\right)^3 - \left(D_i \sim D_a\right)^3\right]f_i} \qquad \text{Eq.( A-1)}$$

where

$$\beta = 1 + \frac{4}{13}\times\left(8\phi_{mo}-1\right)D_a\frac{\sum\left(D_i+D_a\right)^2\left(1-\dfrac{3}{8}\dfrac{D_a}{\left(D_i+D_a\right)}\right)f_i}{\sum\left[D_i^3-\left(D_i \sim D_a\right)^3\right]f_i} \qquad \text{Eq. (A-2)}$$

$$\text{and } D_a = \sum D_i f_i$$

In the above equations, $\phi_{mo}$ is the maximum packing of uniform spheres for which a value of 0.61 was used in this study. $D_a$ is the average diameter of particles, $D_i$ is the diameter of the $i$th size fraction. $(D_i \sim D_a)$ is defined as follows [77]:

$$(D_i \sim D_a) = 0 \; for \, D_i \leq D_a \; and$$

$$(D_i \sim D_a) = D_i - D_a \; for \, D_i > D_a$$

The number fraction of the $i$th component, $f_i$ is defined by:

$$f_i = \frac{\left(\upsilon_i / D_i^3\right)}{\sum\left(\upsilon_i / D_i^3\right)} \qquad \text{Eq. (A-3)}$$

where $\upsilon_i$ is the volume fraction of the $i$th component in the particle population. A maximum packing fraction value, $\phi_m$=0.86 was determined for our particle size distribution assuming that the Dechlorane Plus particles can also be represented as spherical particles using their equivalent diameter values in the analysis.

**Wettability of the particles by the binder:**

The contact angles that the sessile drops of the PDMS made on borosilicate glass and Dechlorane Plus films were measured using a goniometer [109]. The borosilicate glass slides were cleaned via sonication for 10 minutes each in acetone, DI water and isopropyl alcohol, respectively. The slides were then dried in a vacuum oven overnight. The Dechlorane Plus films were prepared by blade casting of a dilute solution of Dechlorane Plus onto the borosilicate glass surface and then drying repeatedly until a smooth and uniform layer of Dechlorane Plus was obtained. The films were dried in a vacuum oven overnight to remove any remaining solvents. A CCD (charge-coupled-device) based camera, equipped with a Melles-Griot 20X magnification telescope was mounted on a vibration damping table to capture the sessile drops of PDMS (Supplemental file, Fig. S2). The image analysis for the determination of the equilibrium contact angles was performed via ImageJ.

**Cryo-Scanning Electron Microscopy**

Upon exit from the capillary die the extrudates were collected in a liquid nitrogen bath at -199°C and then transferred to a cryo-scanning electron microscope. The extrudates were placed into a humidity-free cryo-transfer system (Leica EM VCT100) before SEM imaging. In order to reveal the cross section morphology of the extrudate, the strand was mounted perpendicularly onto the sample holder and fractured at -130°C. A sublimation process was carried out at -80°C under a vacuum level of $10^{-7}$ mbar for 10 minutes to remove any ice condensed onto the surface during the sample preparation, and cooled down to -130°C again for SEM imaging.

SEM analysis was carried out using a Zeiss Auriga Dual-Beam FIB-SEM (Carl Zeiss Microscopy) at an accelerating voltage of 5 kV using an Everhart-Thornley secondary electron detector. Energy dispersive X-ray mapping was carried out for the distributions of carbon, silicon and chlorine elements.

**Table I**: The determination of the apparent slip layer thickness, $\delta$, using the experimentally determined slip coefficients, $\beta$, and the shear viscosity of the binder fluid (which exhibits Newtonian behavior at low shear stresses and "power law" behavior at high shear stresses- see Fig. 2).

a. **Steady torsional flow:** $m_b = 4.9\, Pa-s^{n_b}$, $s_b = 1/n_b = 1$

| $\phi$ | $\beta, 10^{-7}\, m/(s-Pa^{s_b})$ | $\delta, 10^{-6}\, m$ |
|--------|-----------------------------------|-----------------------|
| 0.62   | 4.8                               | 2.4                   |
| 0.72   | 2.3                               | 1.1                   |
| 0.74   | 1.5                               | 0.7                   |
| 0.76   | 1.2                               | 0.5                   |
| **0.78** | **0.6**                         | **0.3**               |

b. **Capillary flow:** $m_b = 17.9\, Pa-s^{n_b}$, $n_b = 0.8$, $s_b = 1/n_b = 1.25$

| $\phi$ | $\beta, 10^{-7}\, m/(s-Pa^{s_b})$ | $\delta, 10^{-6}\, m$ |
|--------|-----------------------------------|-----------------------|
| 0.62   | 0.7                               | 2.6                   |
| 0.72   | 0.36                              | 1.3                   |
| 0.74   | 0.32                              | 1.2                   |
| 0.76   | 0.27                              | 1.0                   |
| 0.78   | 0.2                               | 0.7                   |

**Table II:** Parameters of the Herschel-Bulkley type viscoplastic equation for different $\phi$.

| $\phi$ | $\tau_0, Pa$ | $m, Pa - s^n$ | $n$ |
|--------|--------------|----------------|-----|
| 0.62 | 3.4 | 163 | 0.8 |
| 0.72 | 25 | 680 | 0.8 |
| 0.74 | 43 | 914 | 0.8 |
| 0.76 | 82 | 2073 | 0.8 |
| 0.78 | 182 | 2116 | 0.8 |

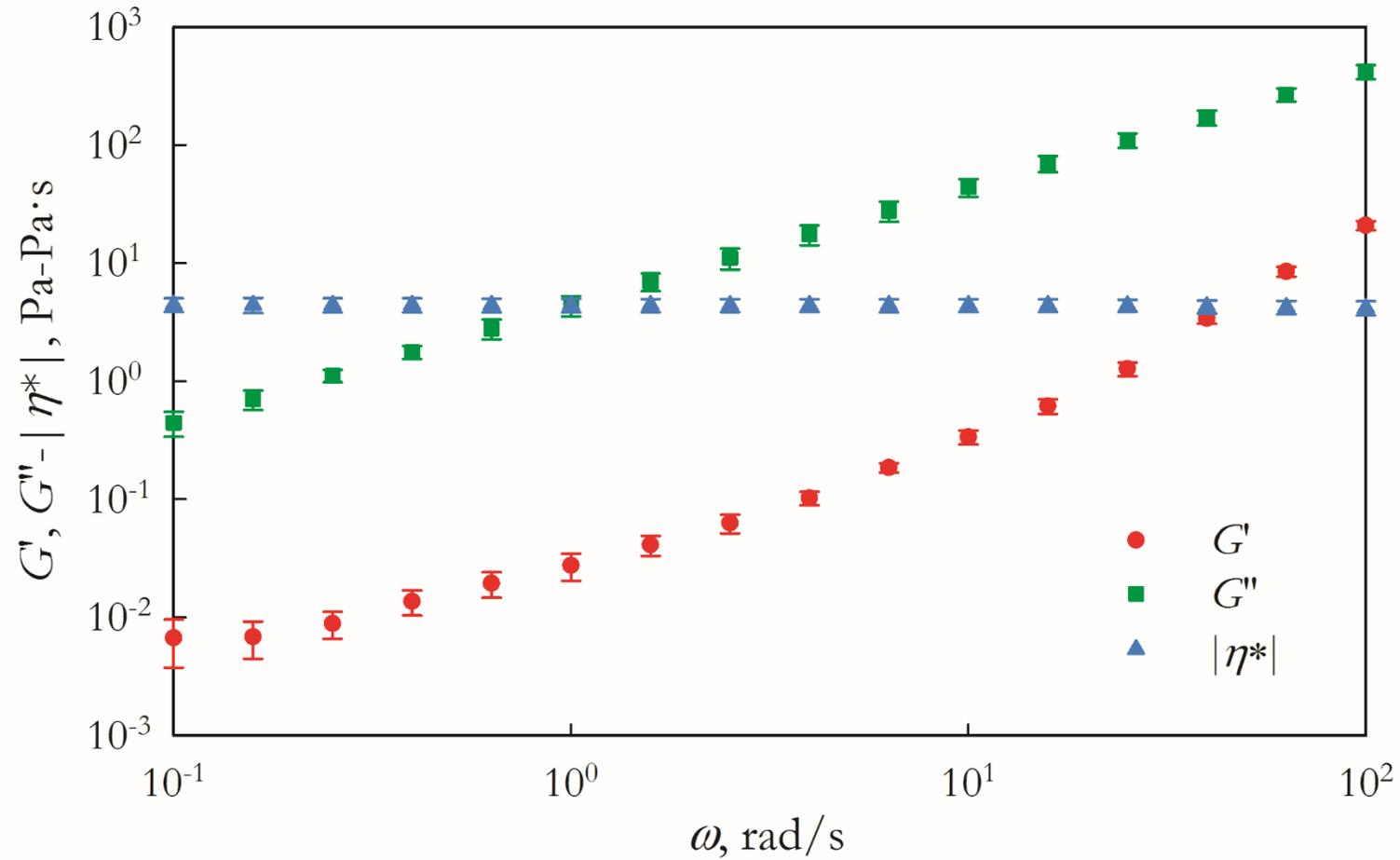

Fig. 1 The storage modulus, $G'$, loss modulus, $G''$, and magnitude of complex viscosity, $|\eta^*|$, values as a function of frequency for the PDMS binder at the strain amplitude of 100%.

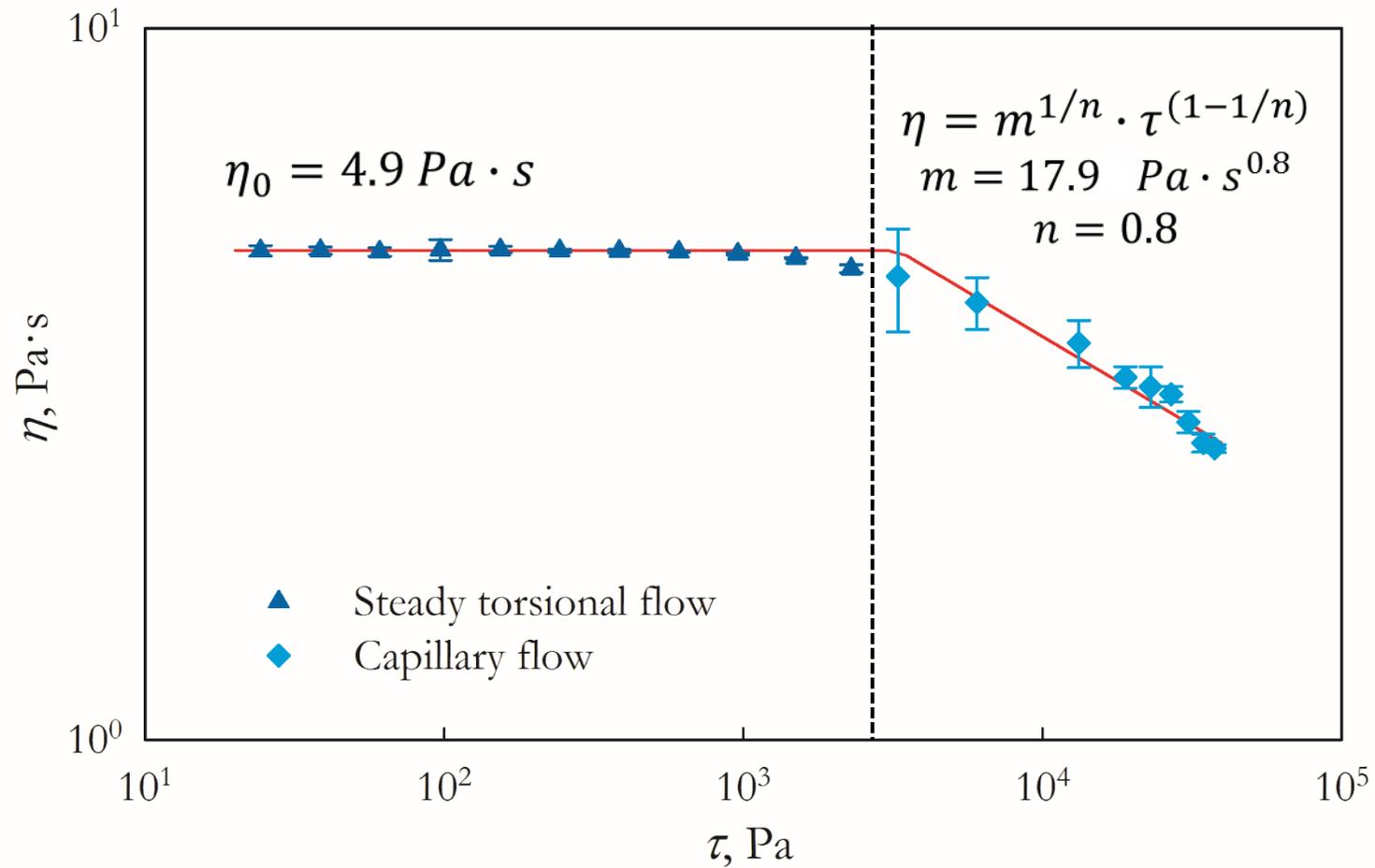

Fig. 2 Shear viscosity of the binder, $\eta$, versus shear stress, $\tau$ obtained from steady torsional flow in the lower shear stress region, and capillary flow in the higher shear stress region.

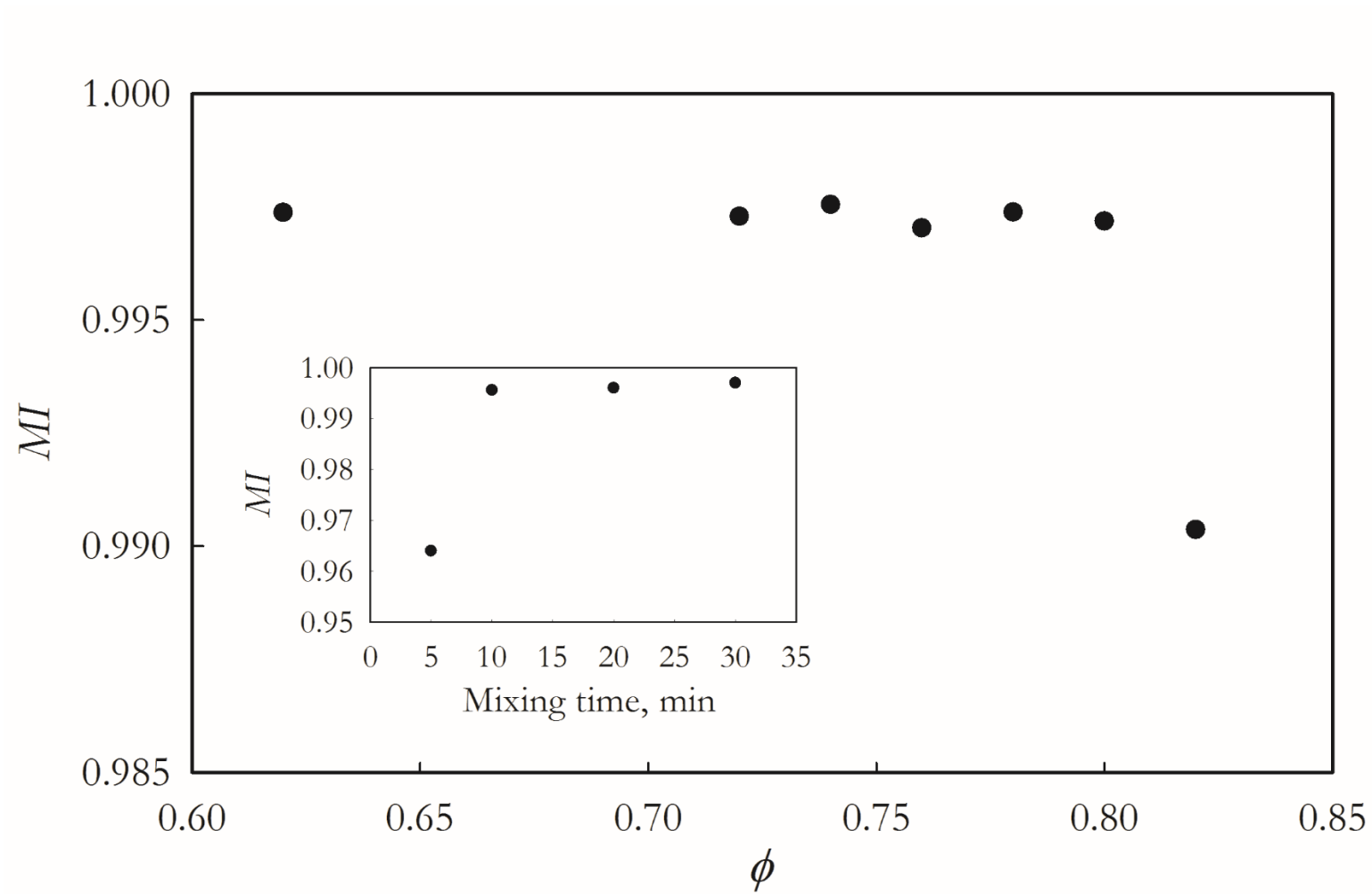

Fig. 3 Mixing index, *MI* versus volume fractions of suspensions. The inset shows the mixing index, *MI* of suspension at volume fraction of $\phi$=0.76 versus mixing time.

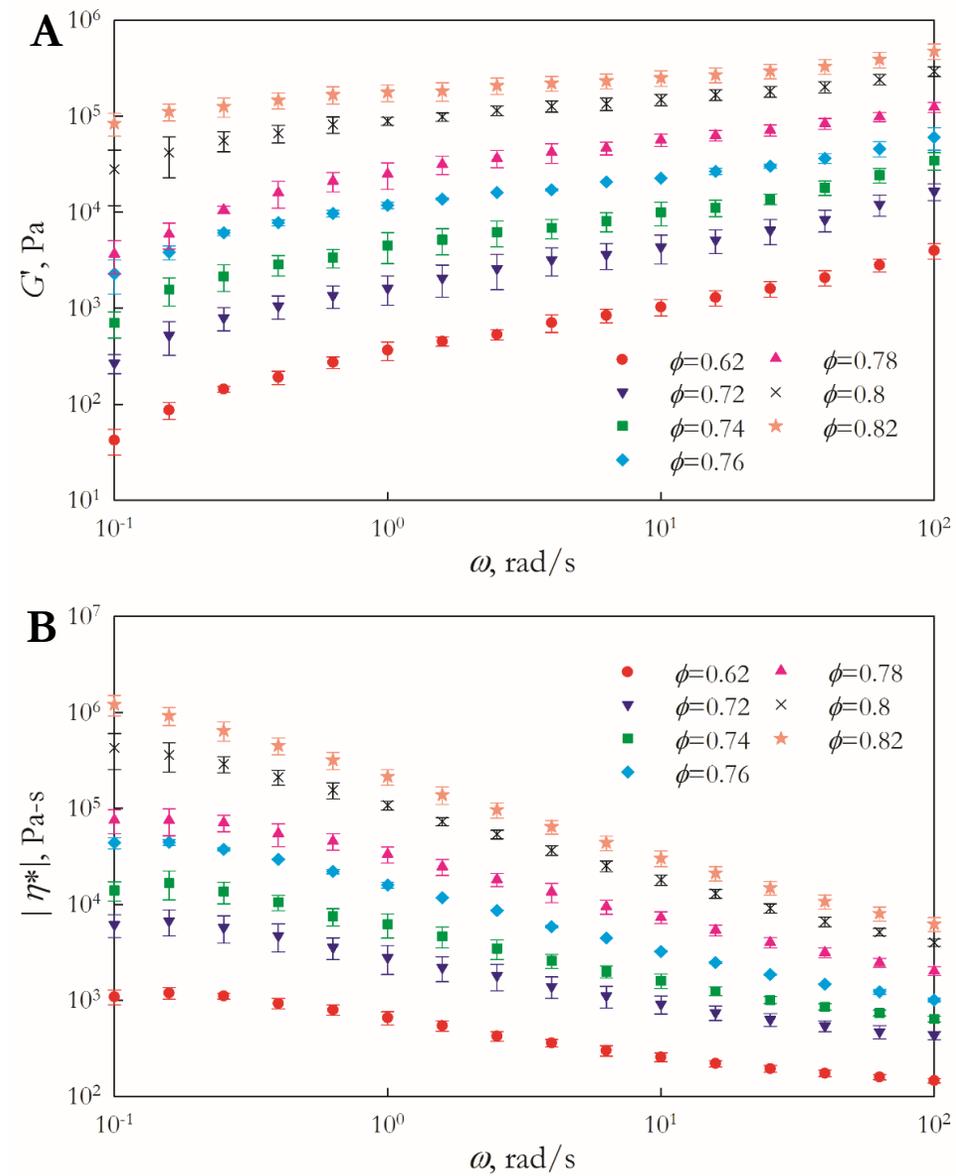

Fig. 4 A) The storage modulus, $G'$, values and B) magnitude of complex viscosity, $|\eta^*|$, as a function of frequency, $\omega$ [4] for the suspensions at the strain amplitude of 0.02%.

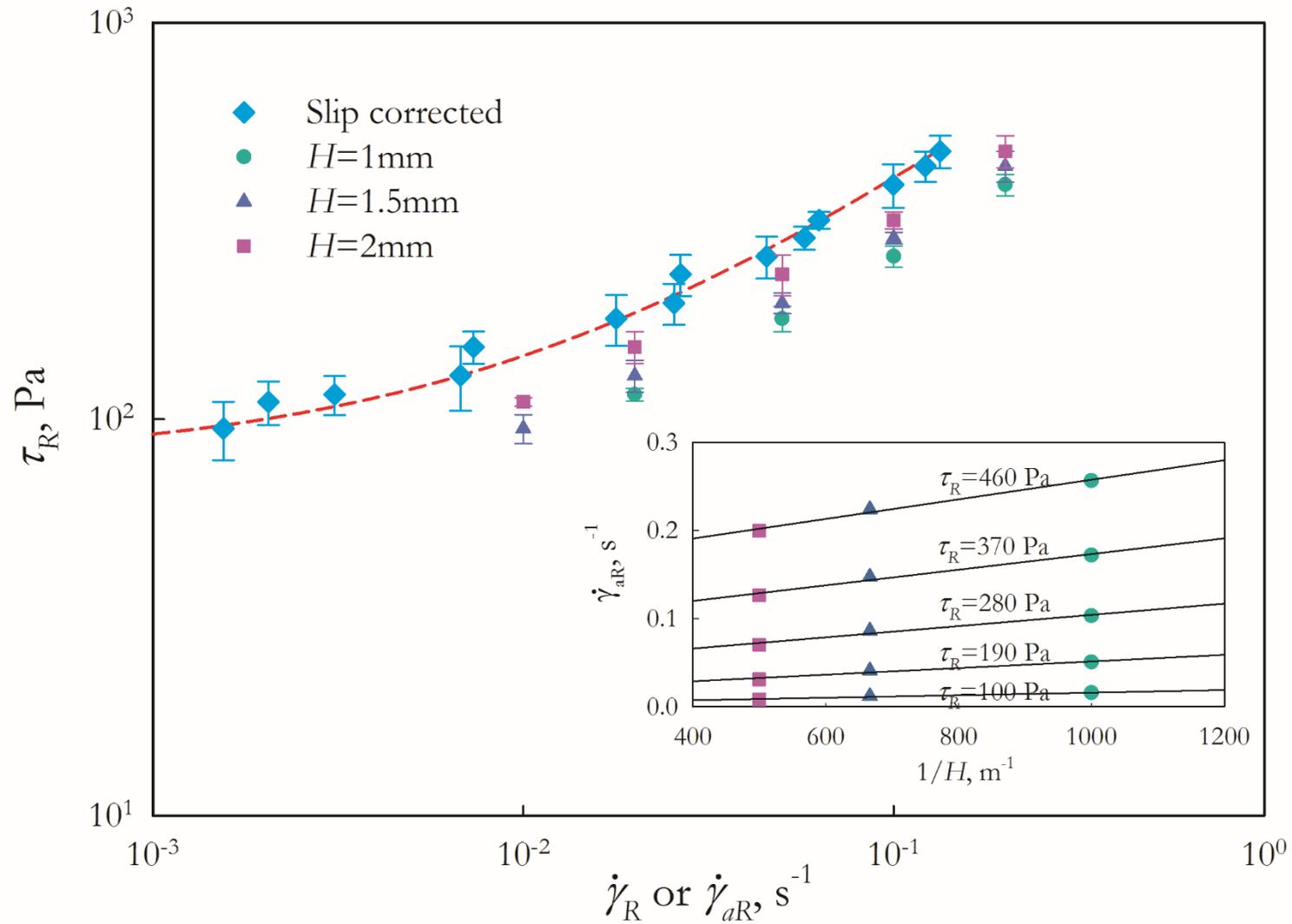

Fig. 5 Shear stress at the edge, $\tau_R$, versus the apparent shear rate at the edge, $\dot\gamma_{aR}$ for different gaps or the slip corrected shear rate at the edge, $\dot\gamma_R$ of $\phi$ =0.76 suspension in the parallel disk torsional flow. The parameters of Herschel-Bulkley equation obtained from the slip corrected data is $\tau_0$=85 Pa, $m$=1436 Pa·s$^{0.67}$, $n$=0.67. Inset shows the apparent shear rate versus reciprocal gap at indicated constant shear stress values in the parallel disk torsional flow tests.

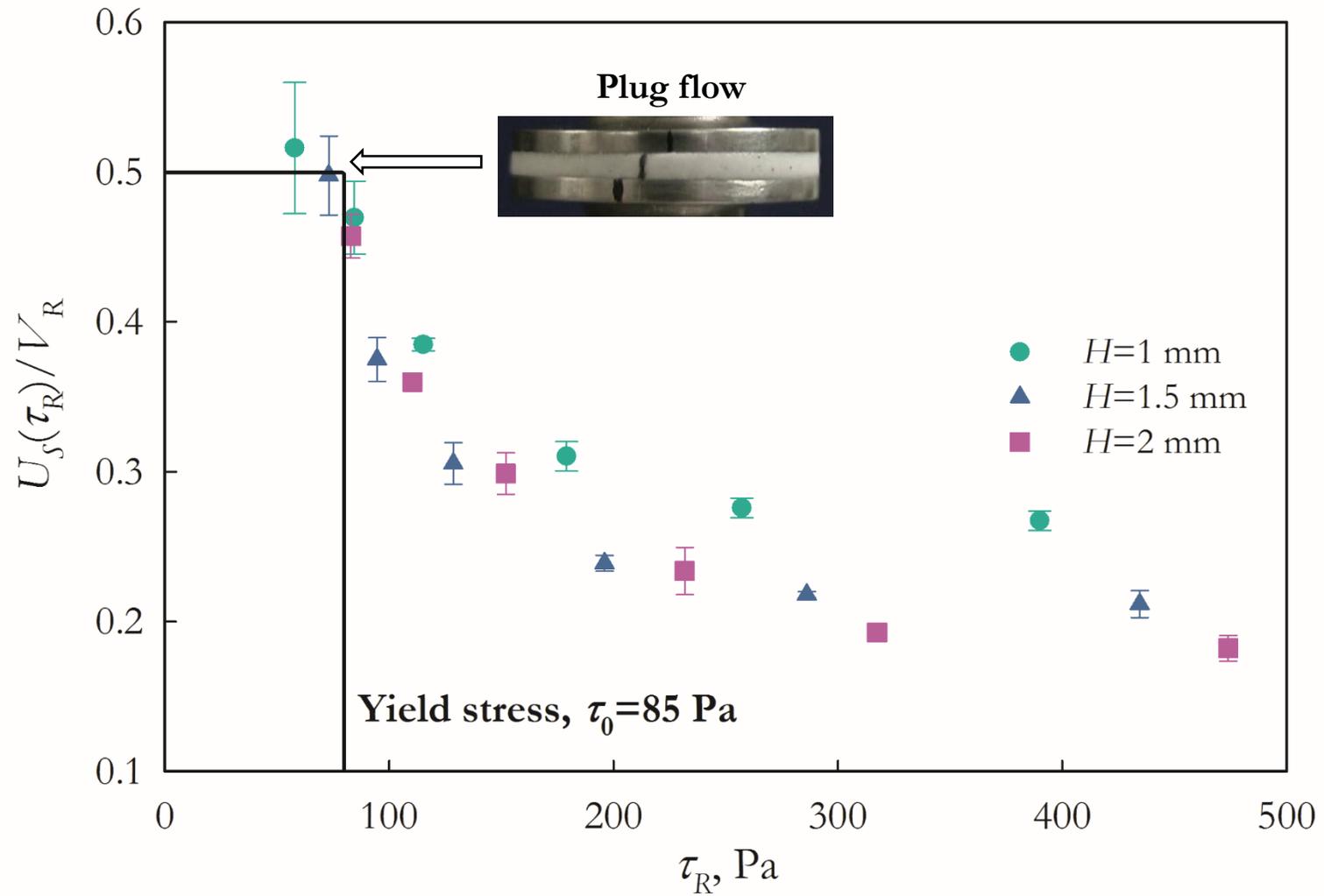

Fig. 6 The ratio of slip velocity over the plate velocity at edge, $U_s(\tau_R)/V_R$, for $\phi$=0.76 at different gaps during steady torsional flow. The inset shows the application of the straight line marker method for plug flow demonstration.

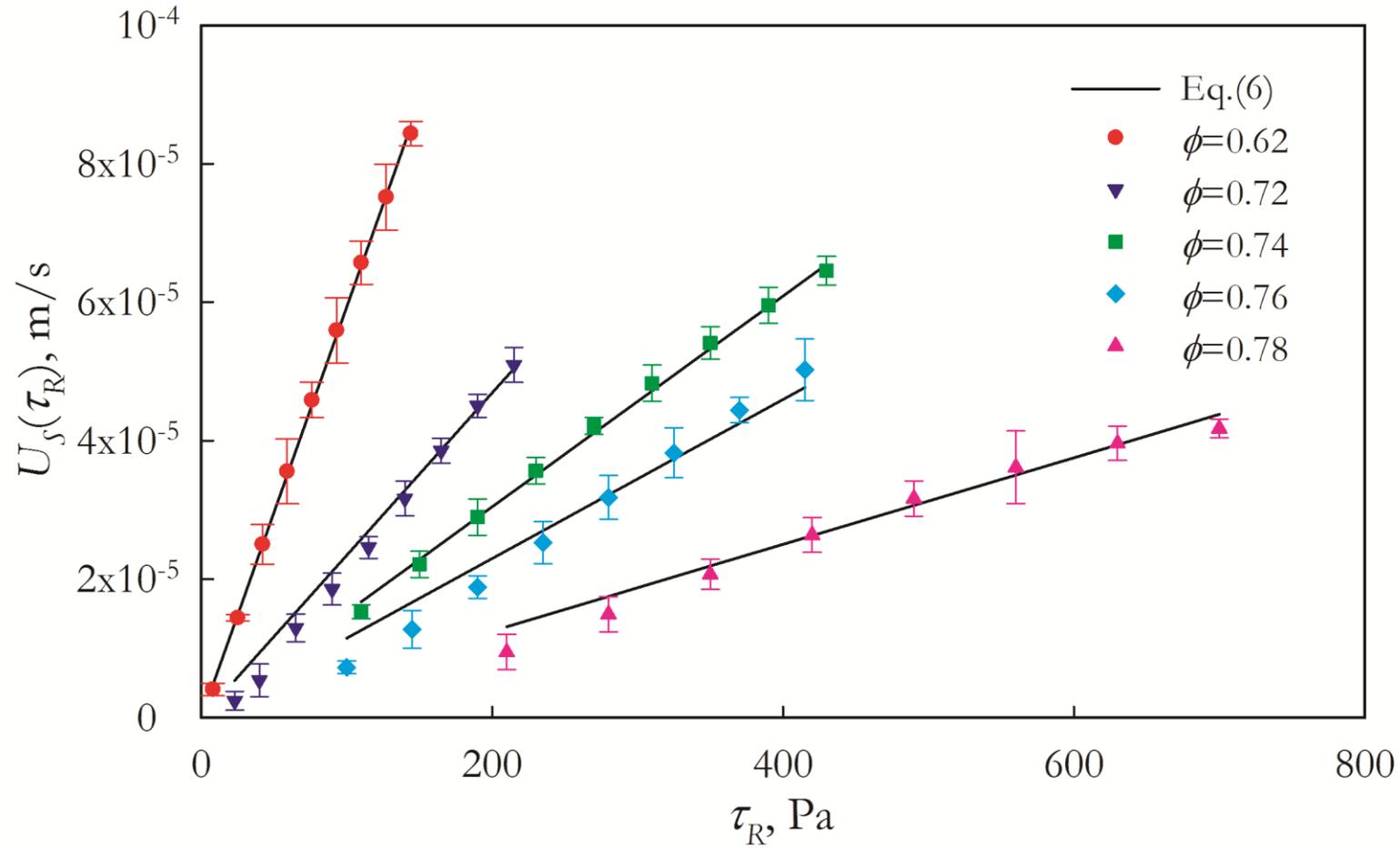

Fig. 7 Slip velocity, $U_S(\tau_R)$, versus shear stress at the edge, $\tau_R$, of suspensions with different volume fractions in steady torsional flow.

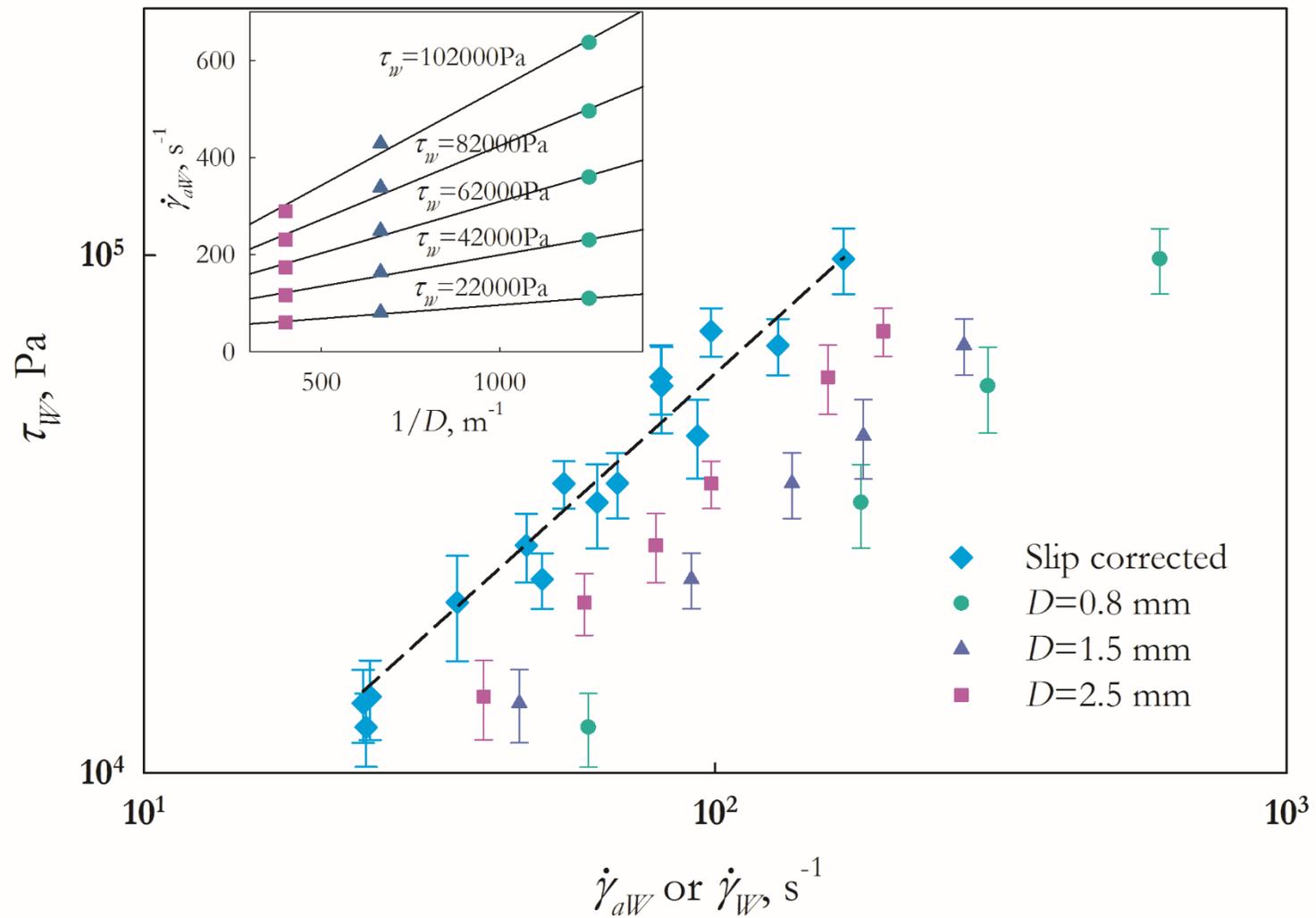

Fig. 8 Bagley corrected wall shear stress, $\tau_W$, versus apparent wall shear rate, $\dot{\gamma}_{aW}$ or slip corrected shear rate, $\dot{\gamma}_W$ in capillary flow of $\phi=0.76$ trimodal suspension. Inset shows apparent shear rate versus reciprocal die diameter at indicated constant shear stress values.

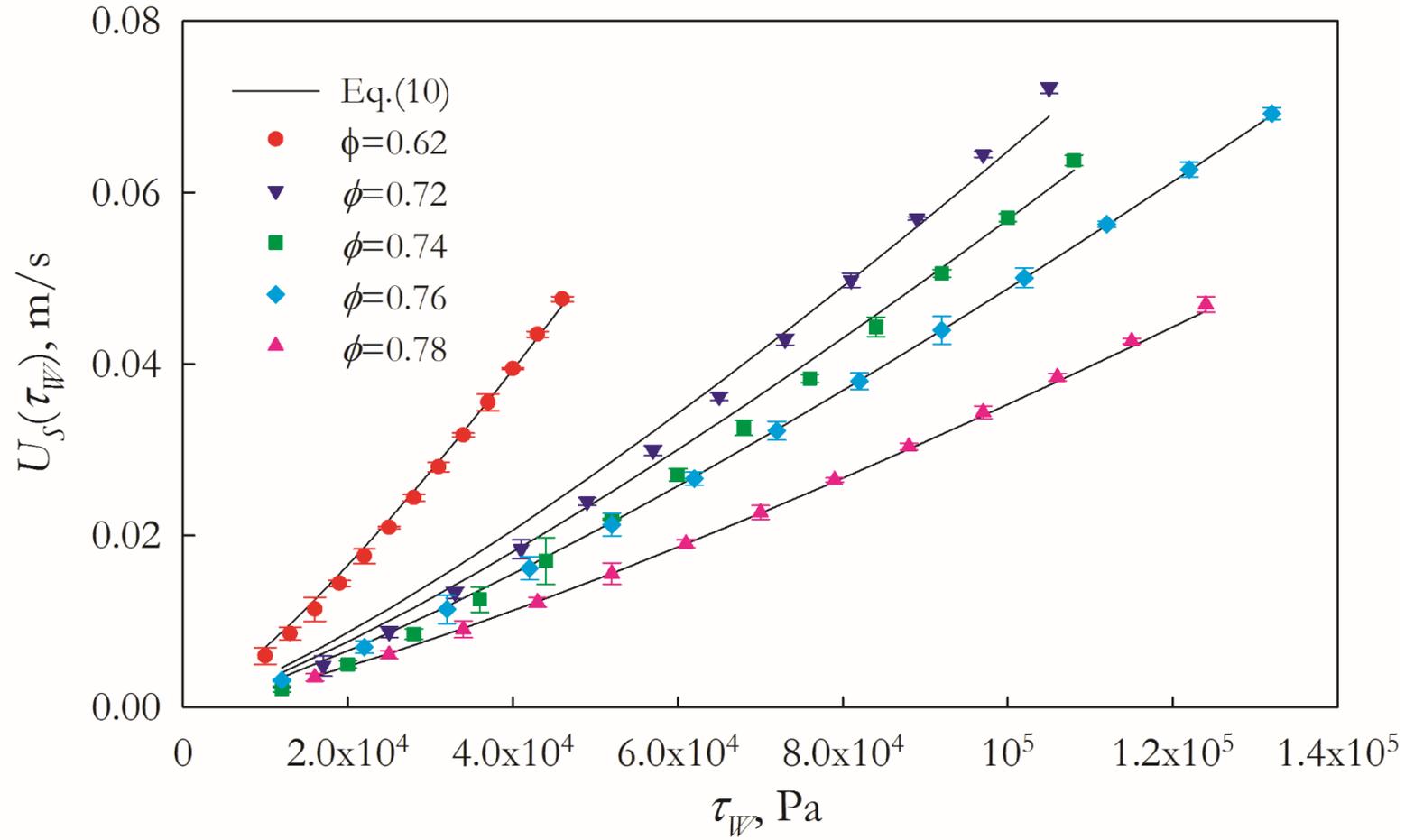

Fig. 9 Slip velocity, $U_S$, versus shear stress at the edge, $\tau_R$, of suspensions with different volume fractions of solids, $\phi$, in capillary flow.

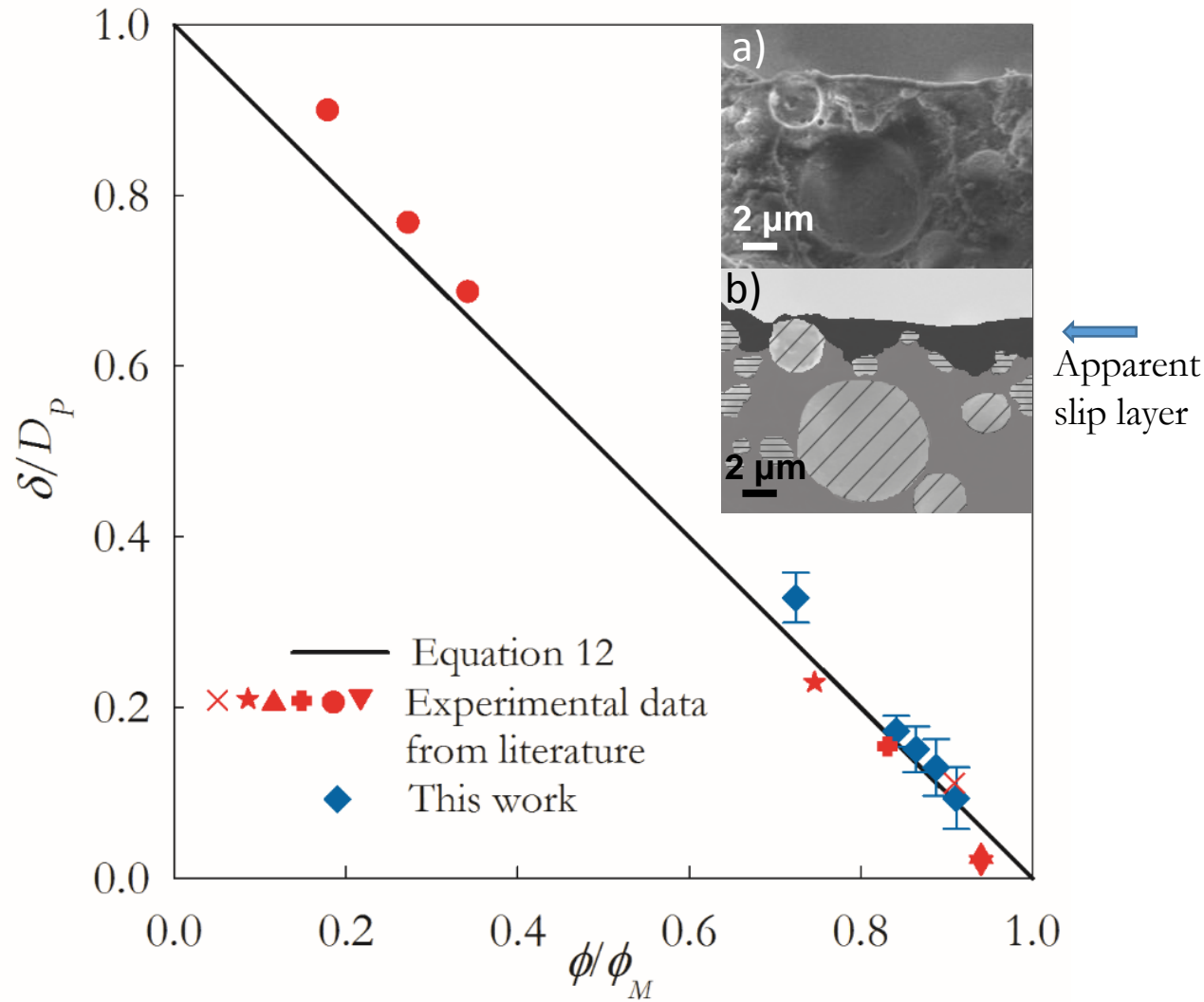

Fig. 10 Correlation of the ratio of apparent slip layer thickness over the harmonic mean particle diameter, $\delta/D_p$, versus the volume fraction over the maximum packing fraction of rigid particles, $\phi/\phi_m$. The data points are experimental results from: ♦ this work, × [Yilmazer and Kalyon (1989)], ★ [Suwardie (1998)], ▲ [Aral and Kalyon (1994)], □ [Kalyon et al. (1995)] ● [Erol et al. (2004)], ▼ [Kalyon et al. (1993)]. The inset shows typical EDX results for the documentation of the apparent slip layer for an extrudate at $\phi$=0.72 .

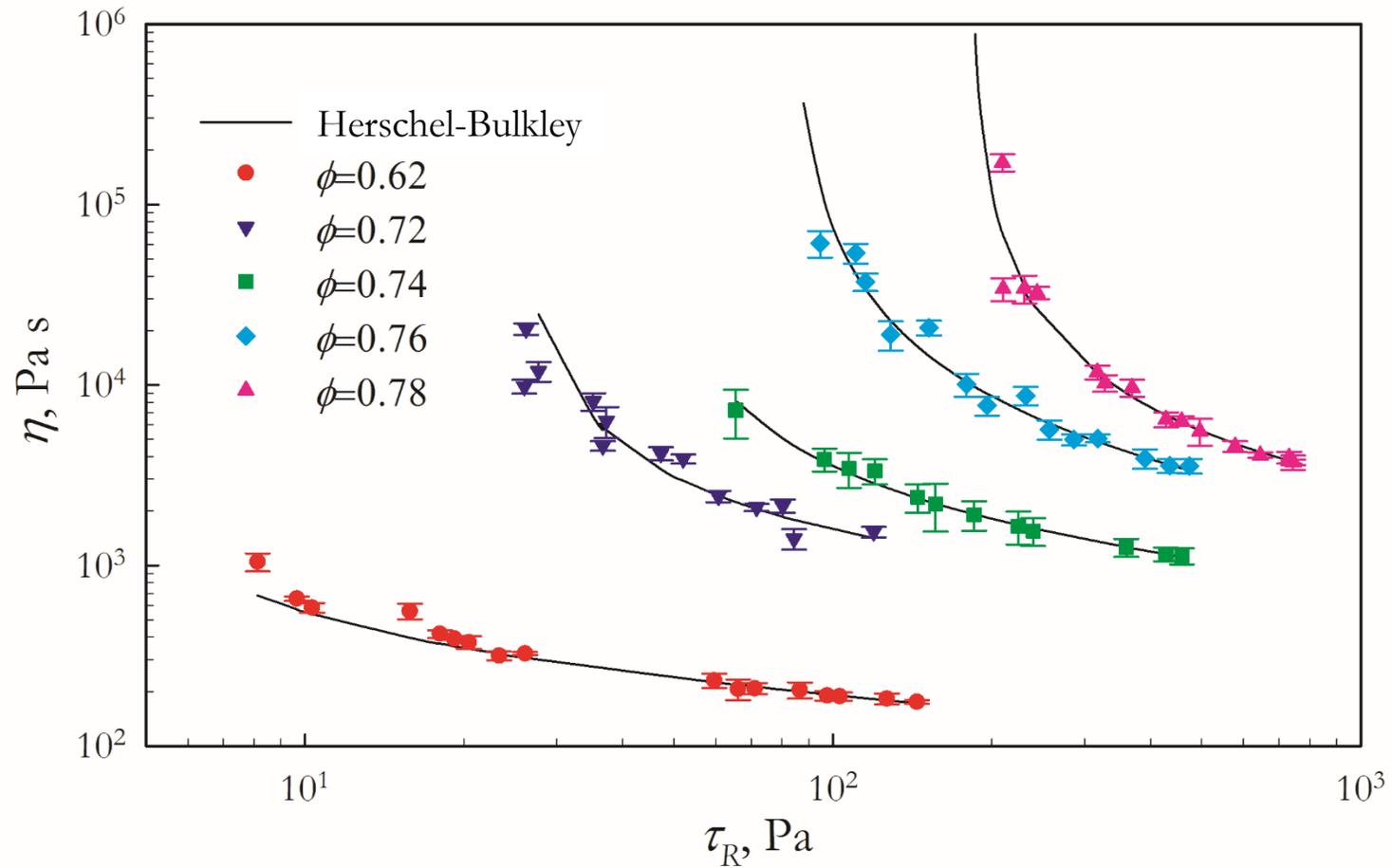

Fig. 11 Wall slip corrected shear viscosity, $\eta$, versus shear stress at the edge, $\tau_R$, of the suspensions obtained from steady torsional flow. The parameters of Herschel-Bulkley equation are given in Table II for all suspensions.

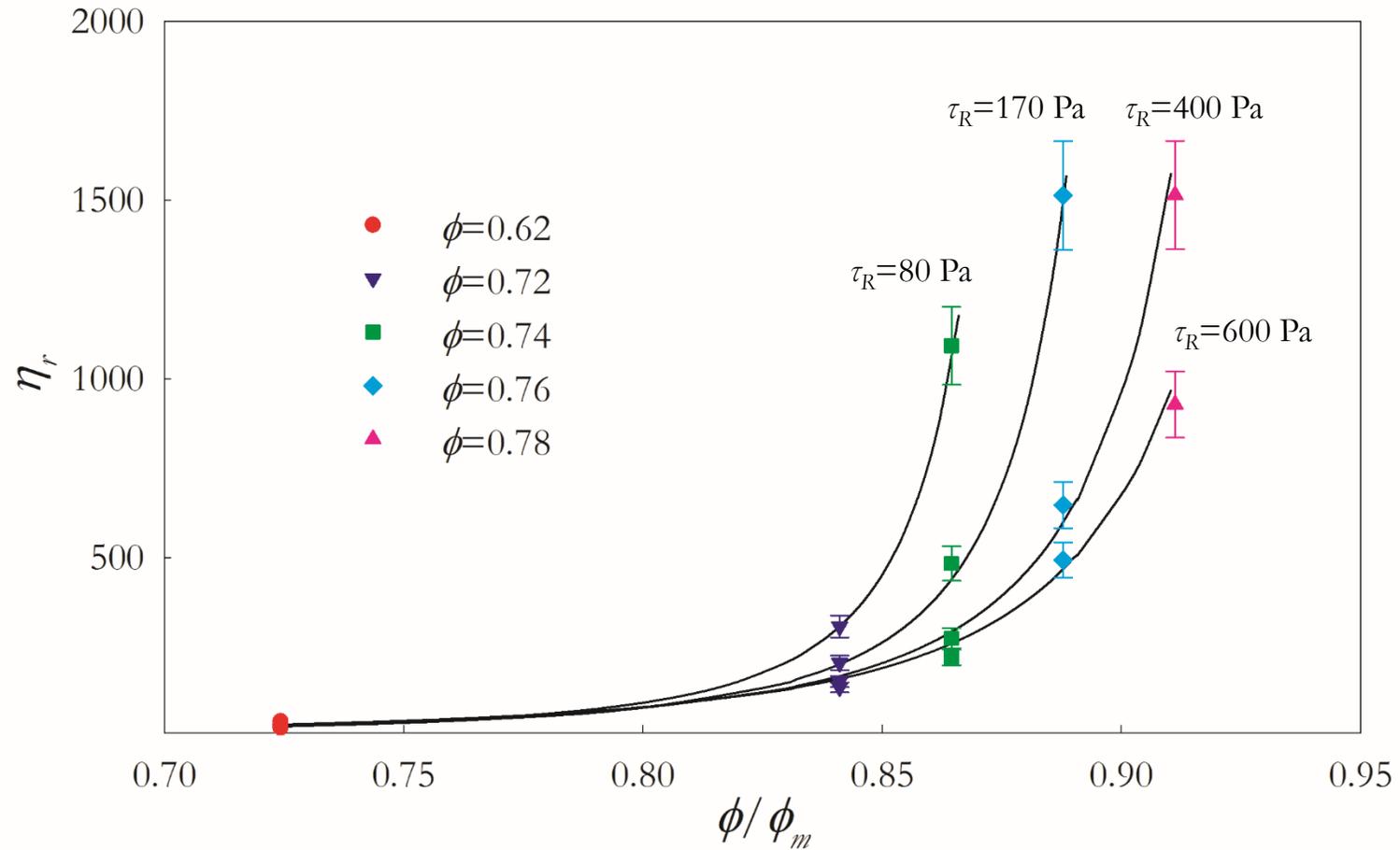

Fig. 12 Wall slip corrected shear viscosity, $\eta$, versus ratio of volume fraction over maximum packing fraction, $\phi/\phi_m$ of suspensions obtained from steady torsional flow at different shear stress values. The fit of the Herschel-Bulkley equation with the parameters given in Table II are shown.

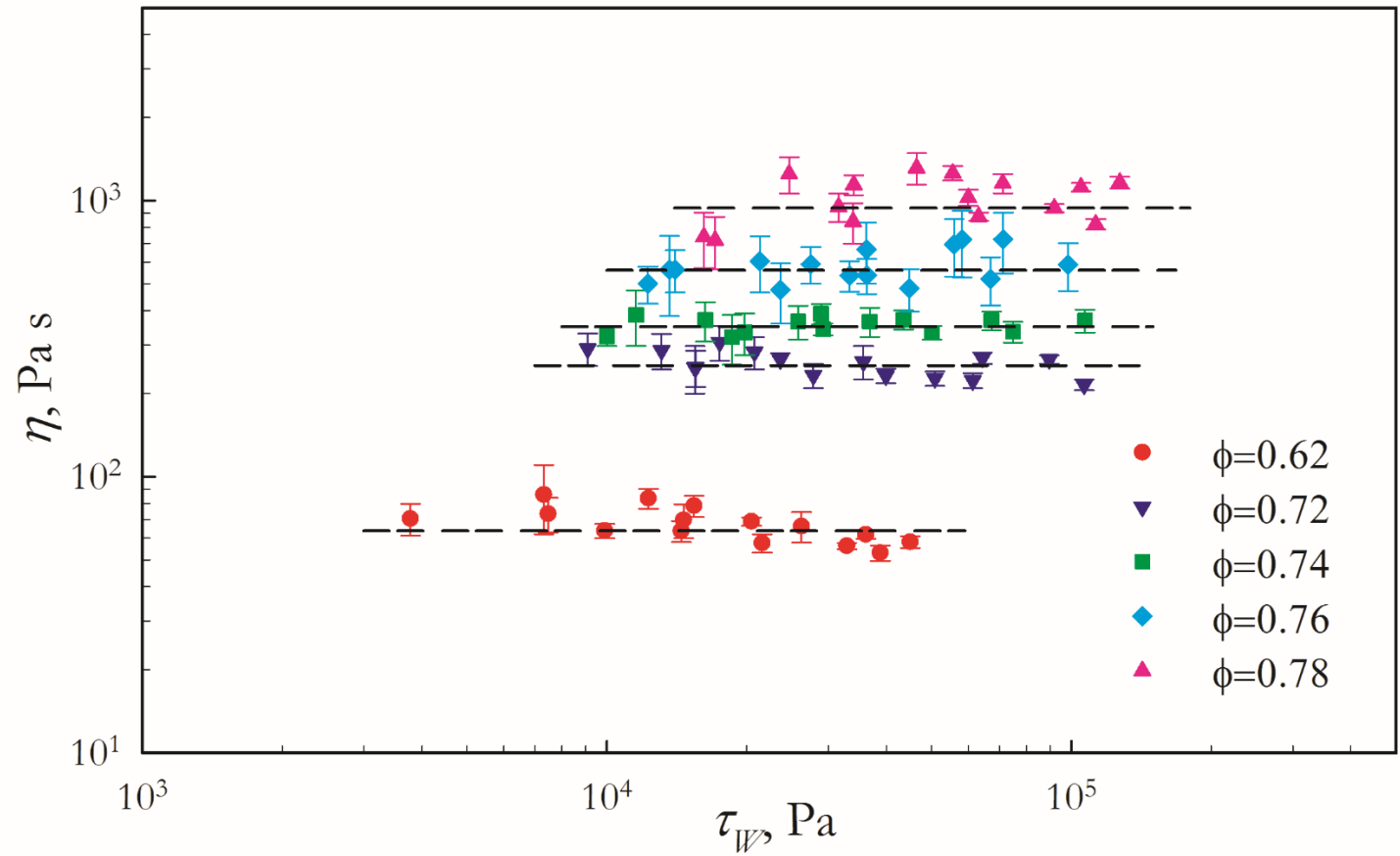

Fig. 13 Wall slip corrected shear viscosity, $\eta$, versus wall shear stress, $\tau_W$, of the suspensions obtained from capillary flow.

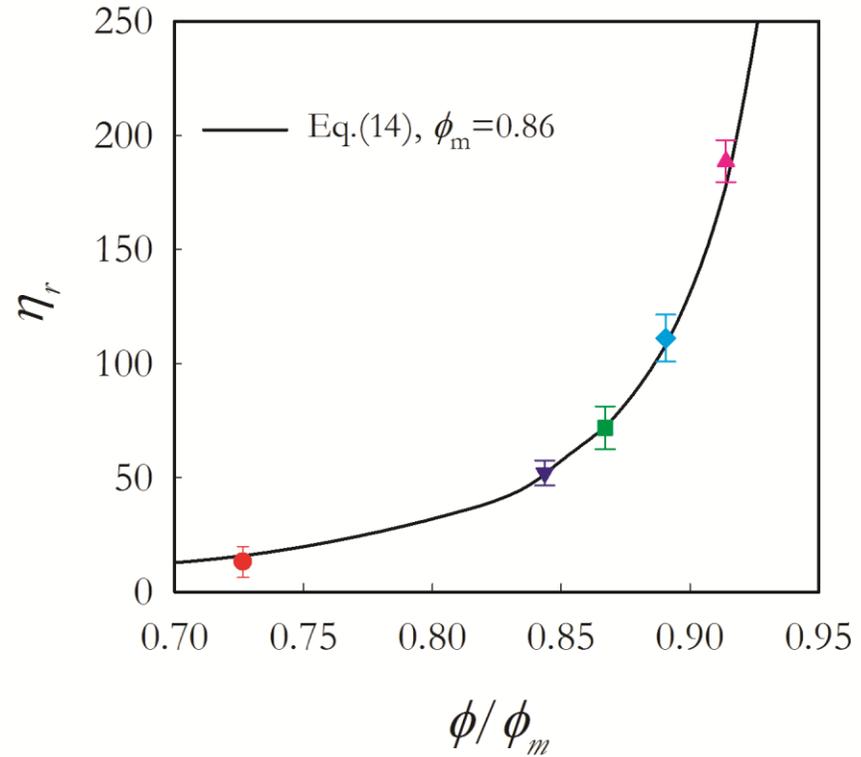

Fig. 14 Relative shear viscosities of the suspensions, $\eta_r$, versus the ratio of volume fraction over maximum packing fraction, $\phi/\phi_m$, at the relatively high wall shear stress values of capillary flow and the prediction of the Krieger-Dougherty equation (Eq. (14)). The maximum packing fraction $\phi_m$=0.86 is determined from the particle size distribution.

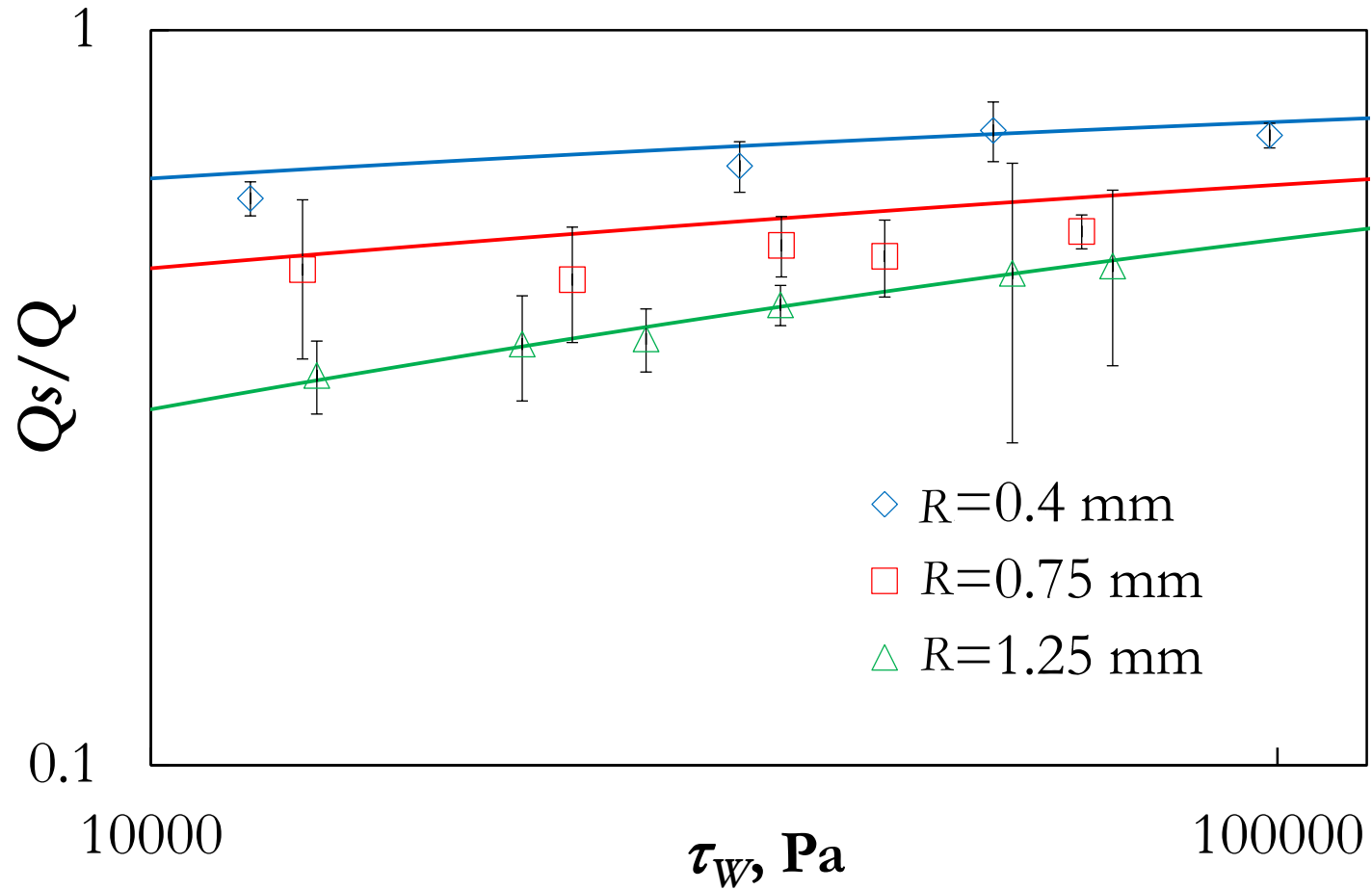

Fig. 15 Ratio of slip volumetric flow rate over total volumetric flow rate, $Q_s/Q$ versus shear stress at the wall, $\tau_W$, for capillary dies with radius of 0.4, 0.75 and 1.25 mm for $\phi$ =0.76. The solid lines are predictions based on Eq. (19) for different capillary radii.

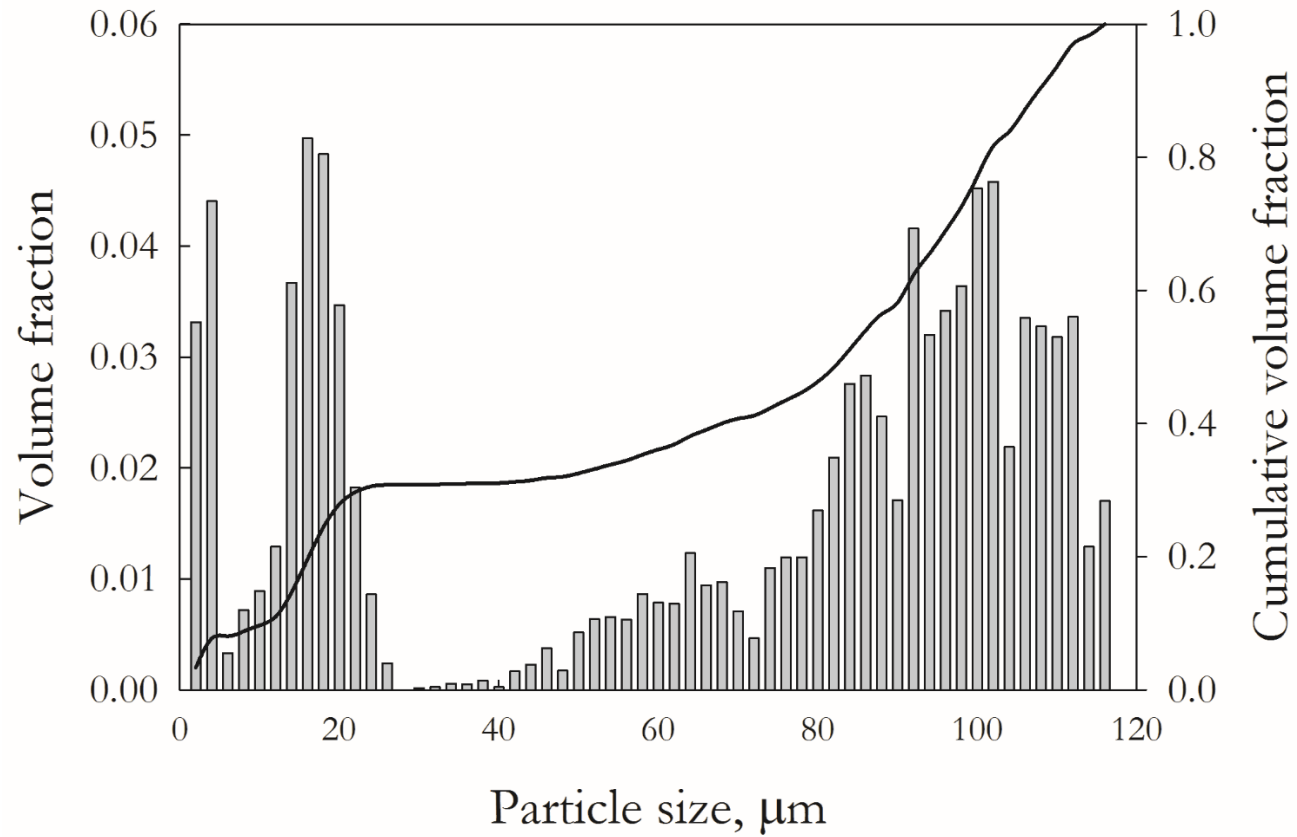

Fig. 16 Volume fraction of each particle size of the trimodal particle mixture(volume ratio: Large:Medium:Small=9:3:1) obtained from the image analysis of the scanning electron microscopy.

# Supplemental figures

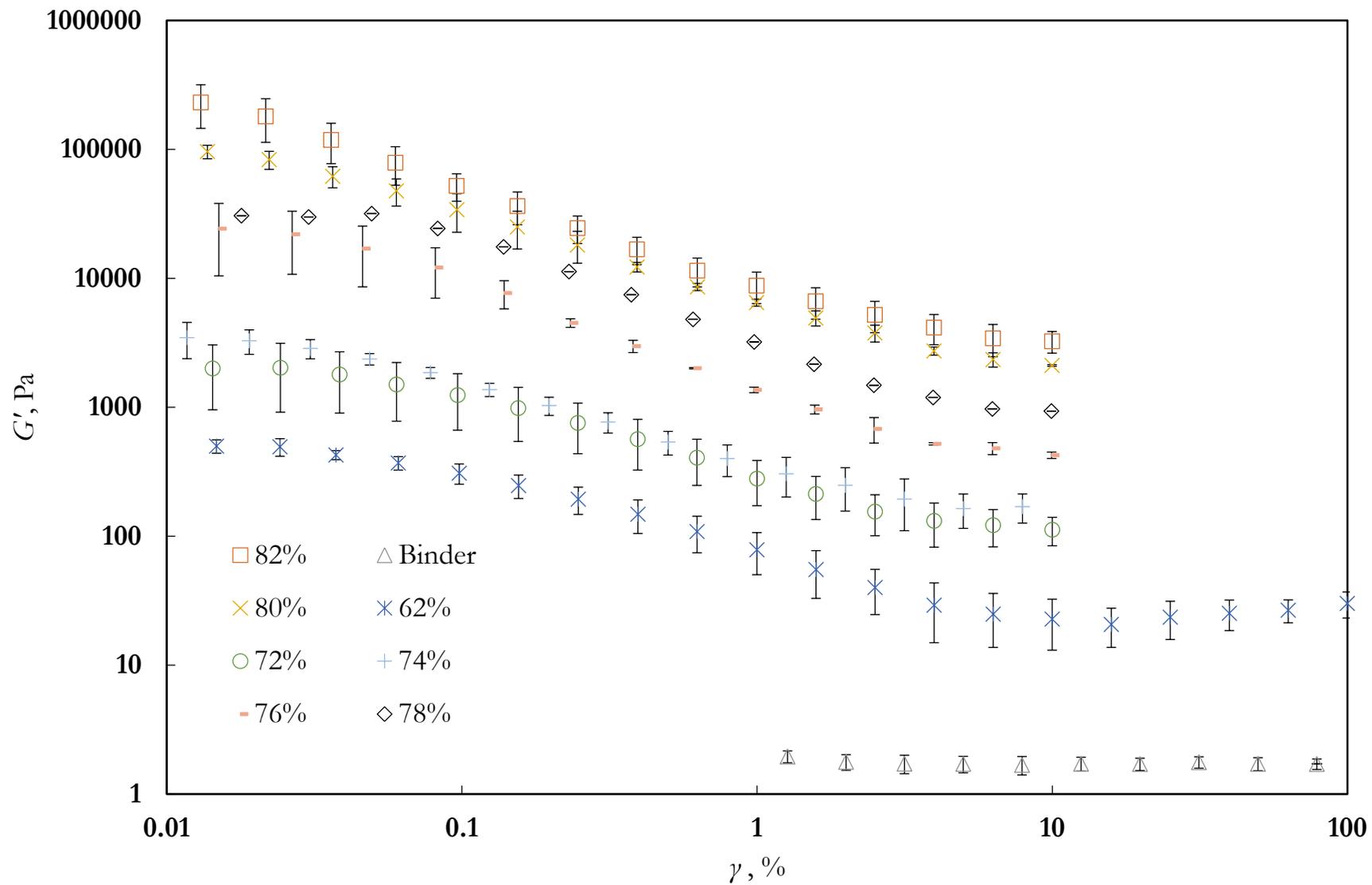

Fig. S1 Storage moduli, G' versus strain amplitude, $\gamma$ of suspensions with different volume fractions and pure binder at a frequency of 1 rad/s

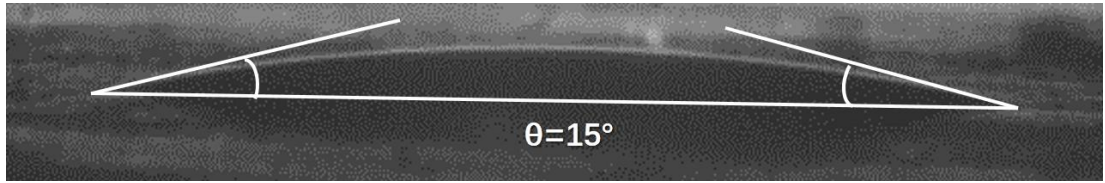

DMS-T35 on Dechlorane plus

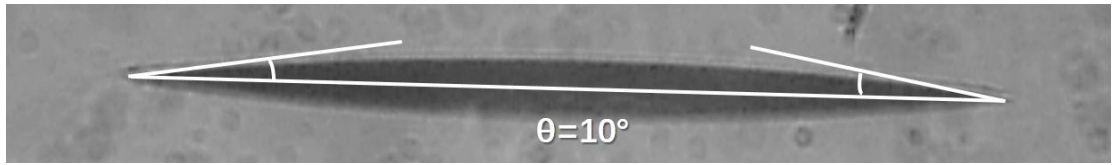

DMS-T35 on Borosilicate glass

Fig. S2 Contact angle of pure binder on substrate of the same construction materials as the packing particles.

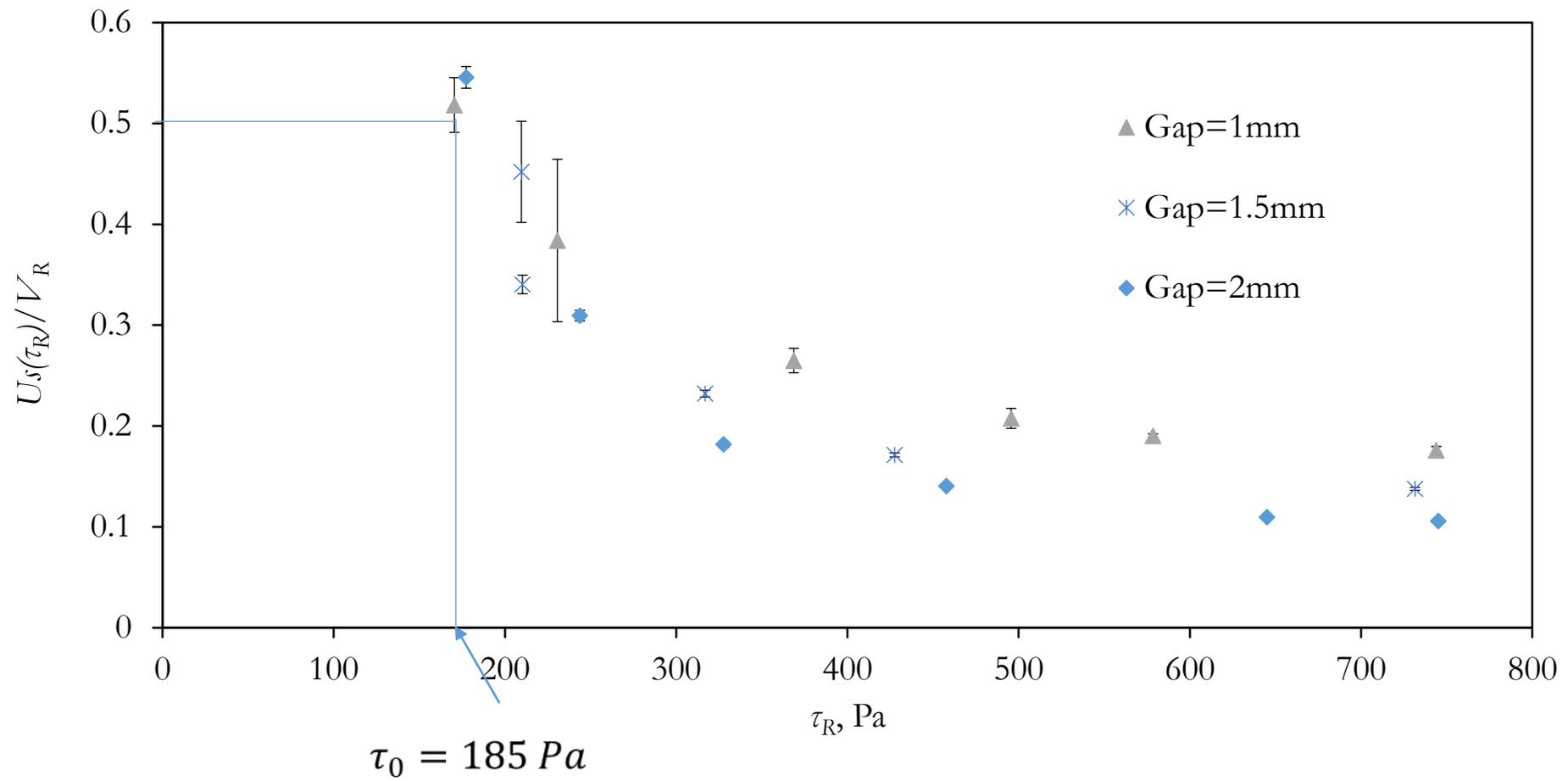

Fig. S3 The ratio of slip velocity over the plate velocity at edge, $U_s(\tau_R)/V_R$, for $\phi$=0.78 at different gaps. The inset shows the application of the straight line marker method.

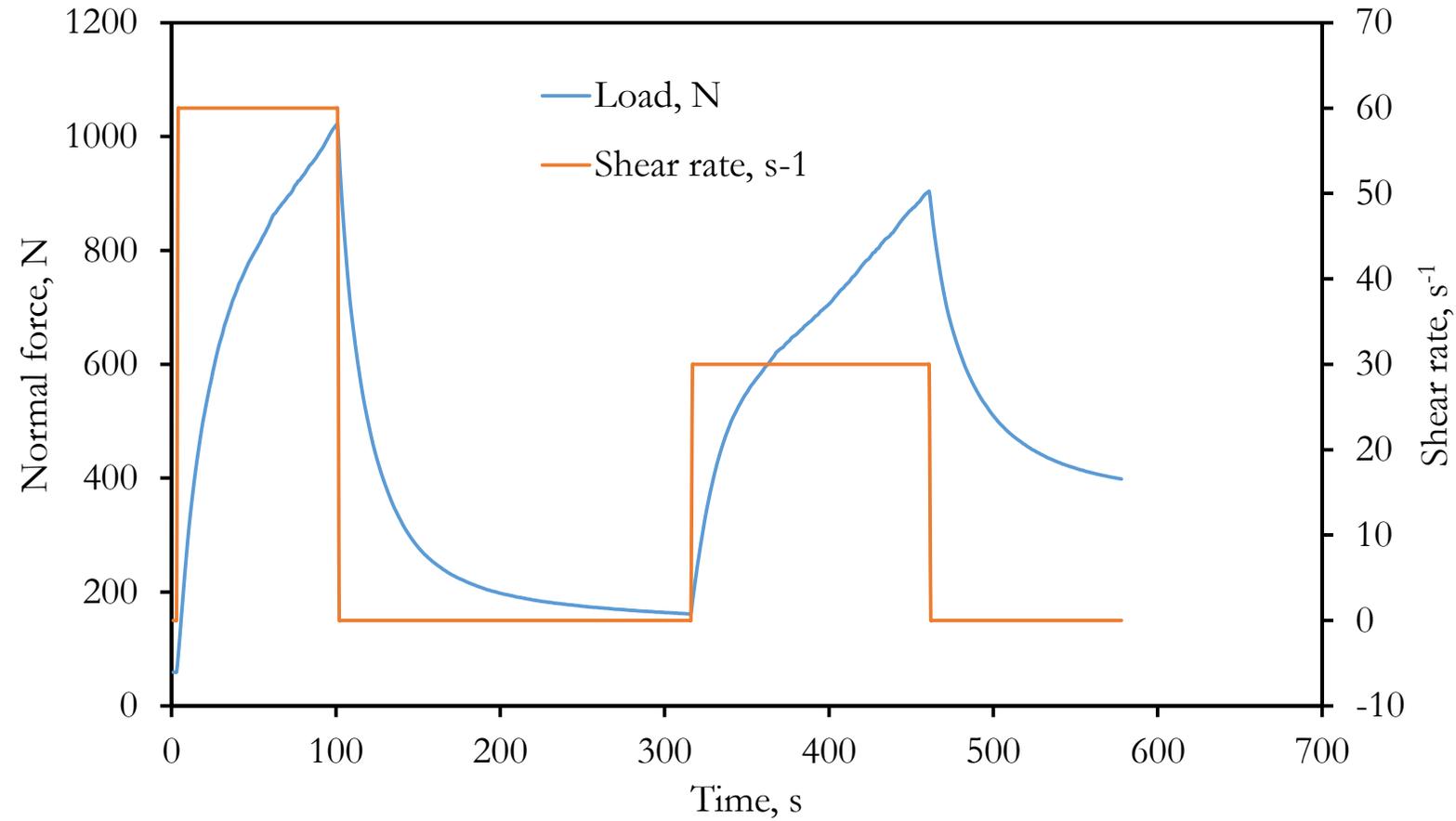

Fig. S4 Normal force and shear rate versus time during capillary test of suspension of $\phi$=0.82. The capillary die has a diameter of 0.8 mm, L/D=20.

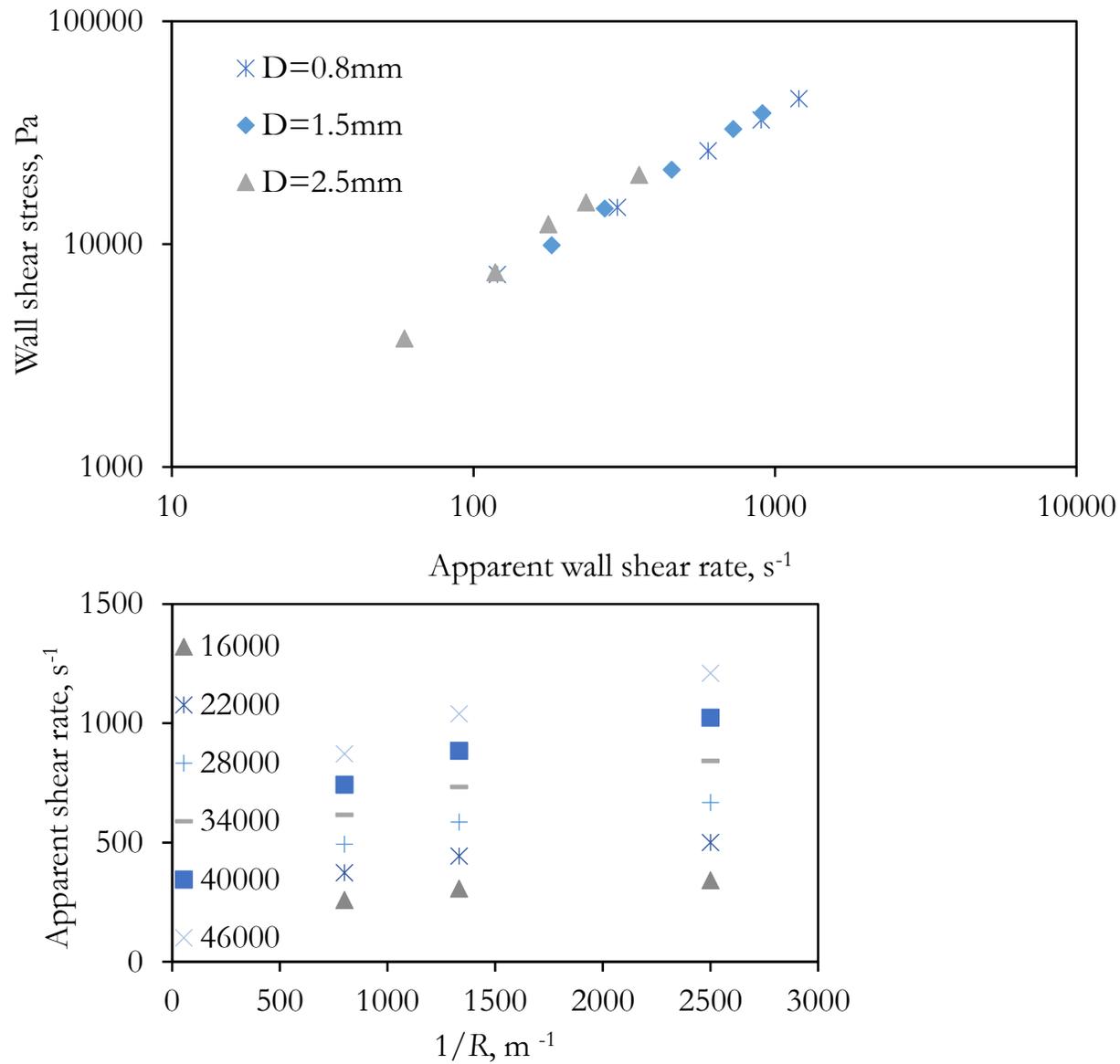

Fig. S5 Upper shows Bagley corrected wall shear stress, $\tau_W$, versus apparent wall shear rate, $\dot{\gamma}_{aW}$ or slip corrected shear rate, $\dot{\gamma}_W$ in capillary flow of $\phi=0.62$ trimodal suspension. bottom shows apparent shear rate versus reciprocal die diameter at indicated constant shear stress values. 22

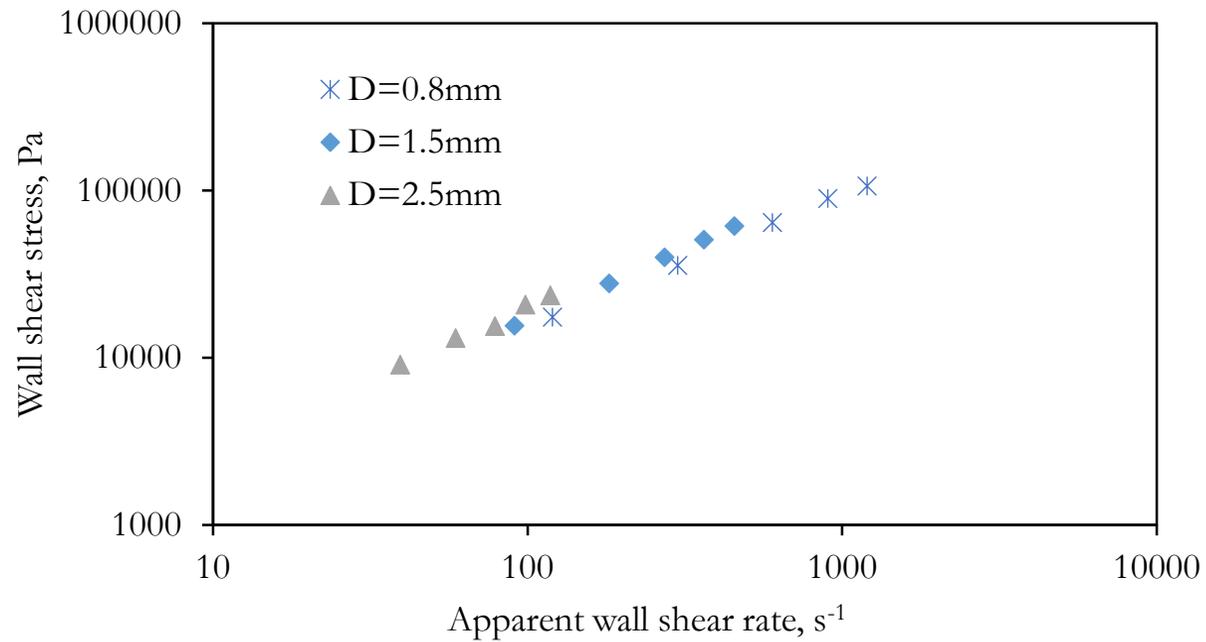

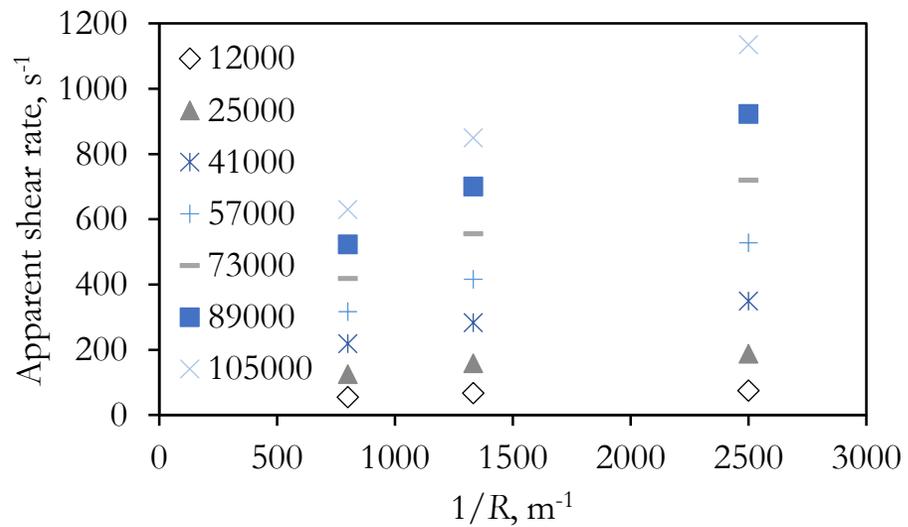

Fig. S6 Upper shows Bagley corrected wall shear stress, $\tau_W$, versus apparent wall shear rate, $\dot{\gamma}_{aW}$ or slip corrected shear rate, $\dot{\gamma}_W$ in capillary flow of $\phi$=0.72 trimodal suspension. bottom shows apparent shear rate versus reciprocal die diameter at indicated constant shear stress values.

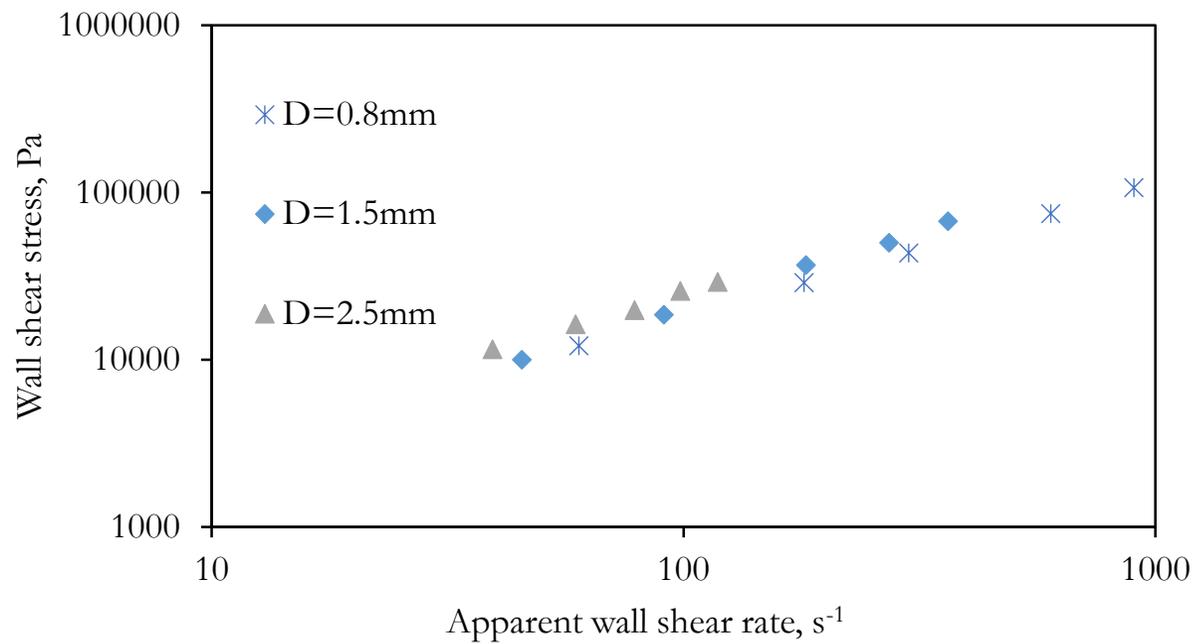

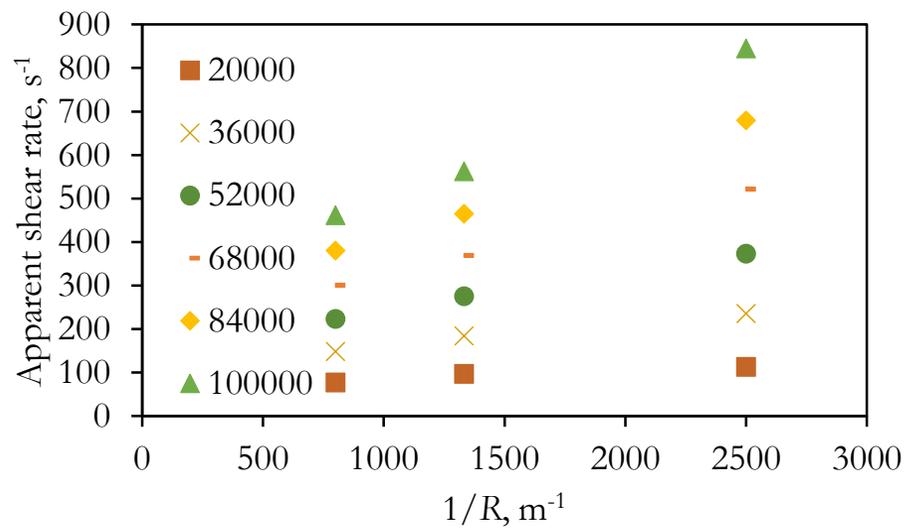

Fig. S7 Upper shows Bagley corrected wall shear stress, $\tau_W$, versus apparent wall shear rate, $\dot{\gamma}_{aW}$ or slip corrected shear rate, $\dot{\gamma}_W$ in capillary flow of $\phi$=0.74 trimodal suspension. bottom shows apparent shear rate versus reciprocal die diameter at indicated constant shear stress values.

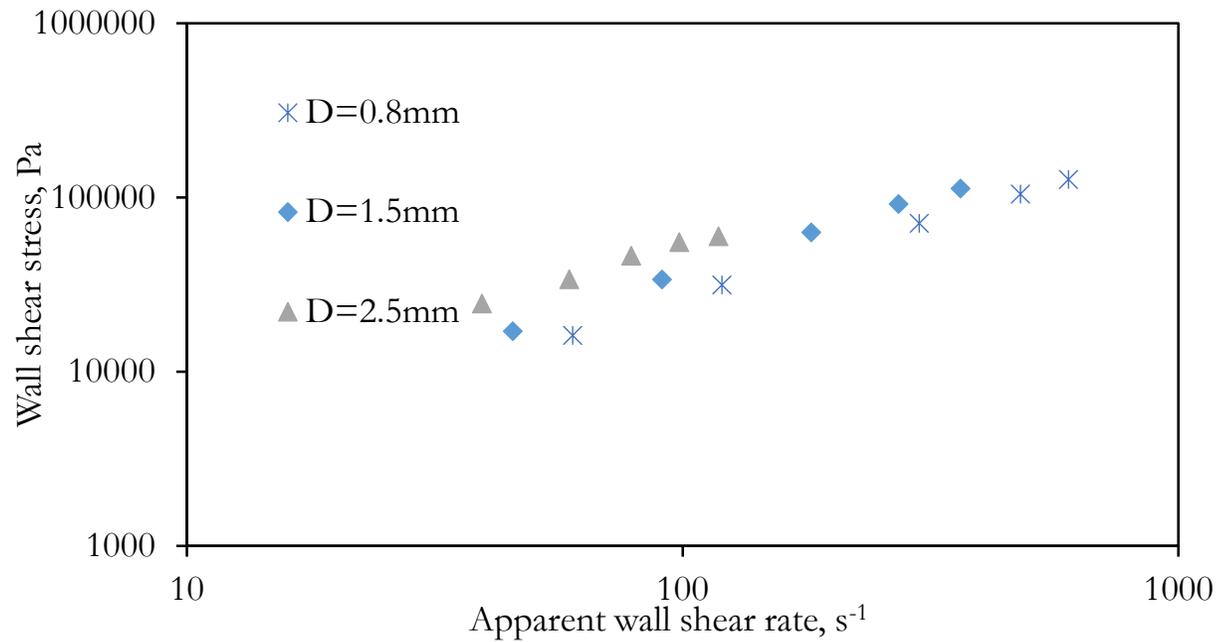

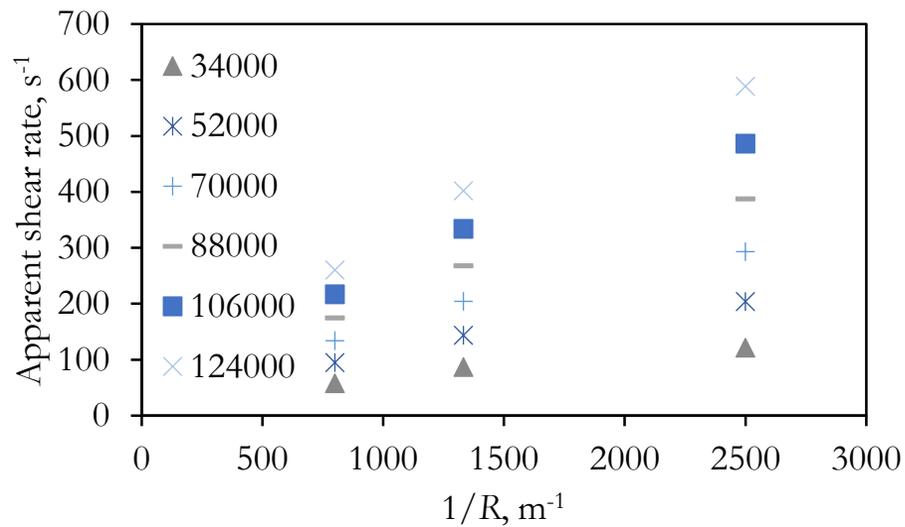

Fig. S8 Upper shows Bagley corrected wall shear stress, $\tau_W$, versus apparent wall shear rate, $\dot{\gamma}_{aW}$ or slip corrected shear rate, $\dot{\gamma}_W$ in capillary flow of $\phi$=0.78 trimodal suspension. Bottom shows apparent shear rate versus reciprocal die diameter at indicated constant shear stress values. 25

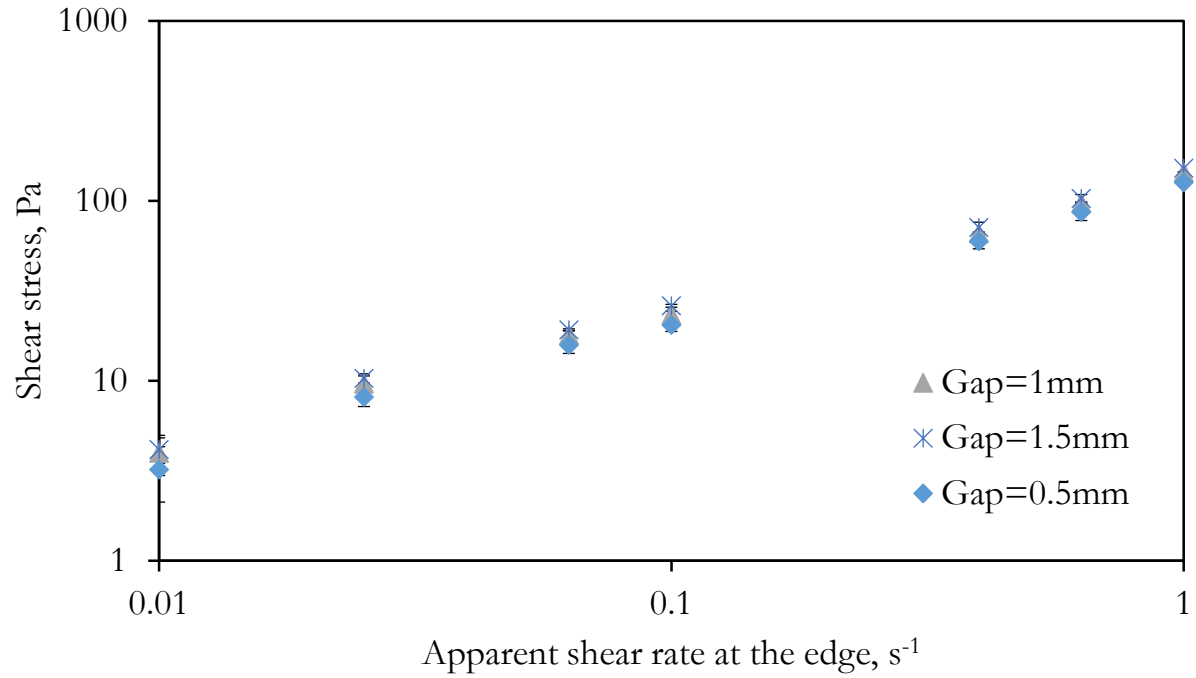

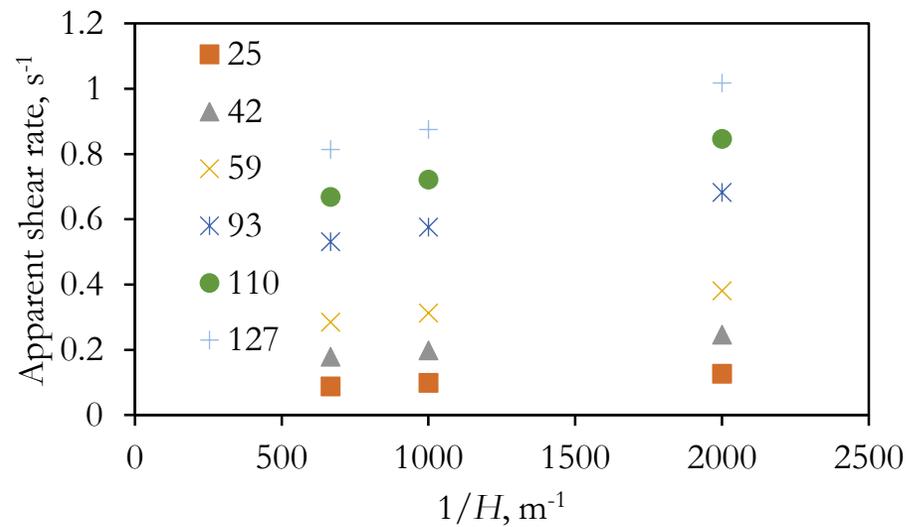

Fig. S9 Upper shows Shear stress at the edge, $\tau_R$, versus the apparent shear rate at the edge, $\dot{\gamma}_{aR}$ for different gaps or the slip corrected shear rate at the edge, $\dot{\gamma}_R$ of $\phi$=0.62 suspension in the parallel disk torsional flow. Bottom shows apparent shear rate versus reciprocal gap at indicated constant shear stress values in the parallel disk torsional flow tests.

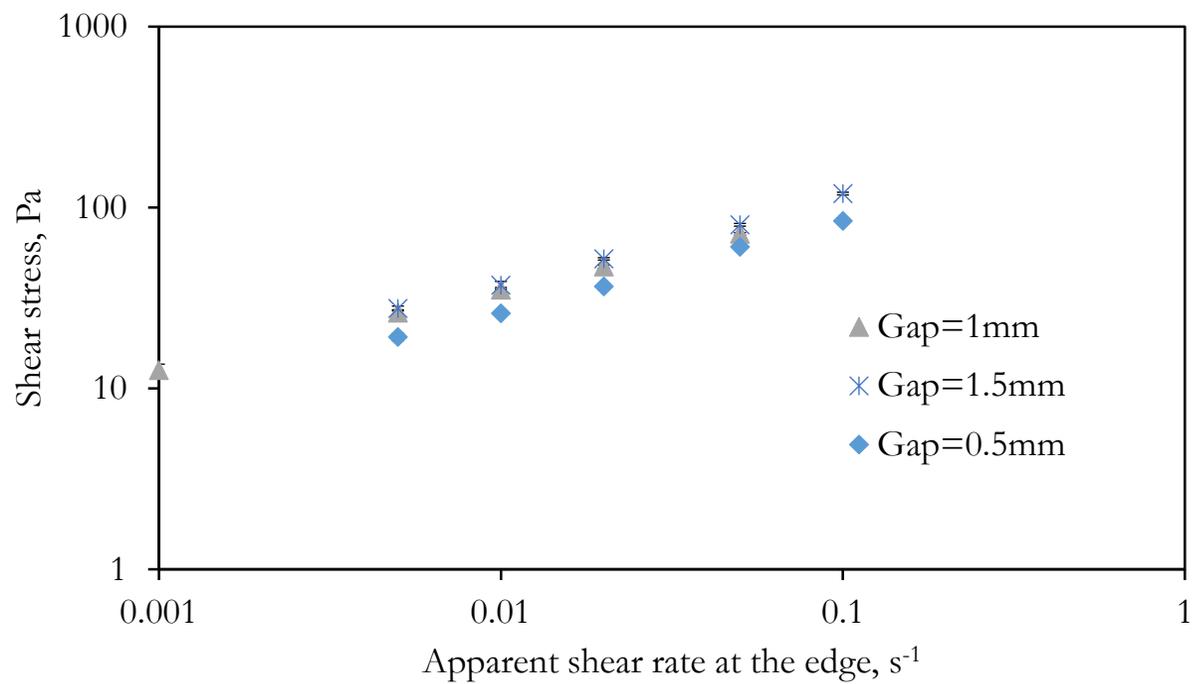

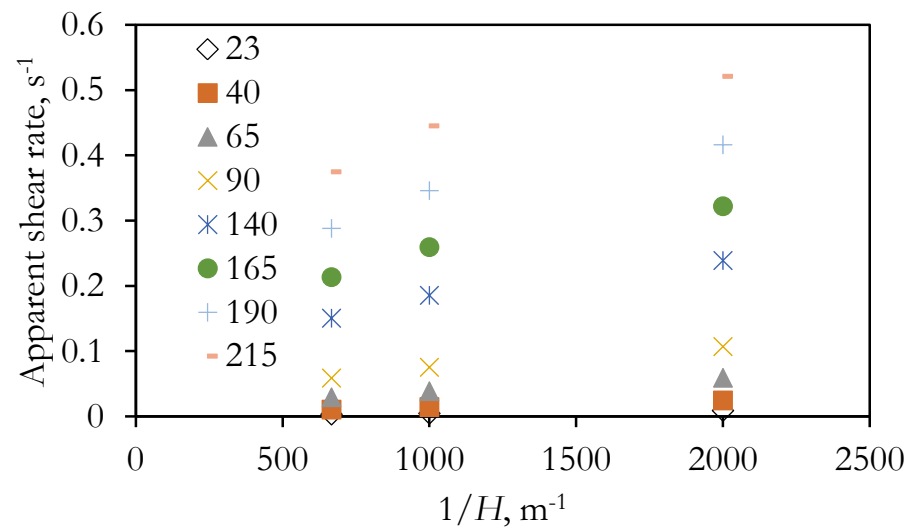

Fig. S10 Upper shows Shear stress at the edge, $\tau_R$, versus the apparent shear rate at the edge, $\dot{\gamma}_{aR}$ for different gaps or the slip corrected shear rate at the edge, $\dot{\gamma}_R$ of $\phi$=0.72 suspension in the parallel disk torsional flow. Bottom shows apparent shear rate versus reciprocal gap at indicated constant shear stress values in the parallel disk torsional flow tests.

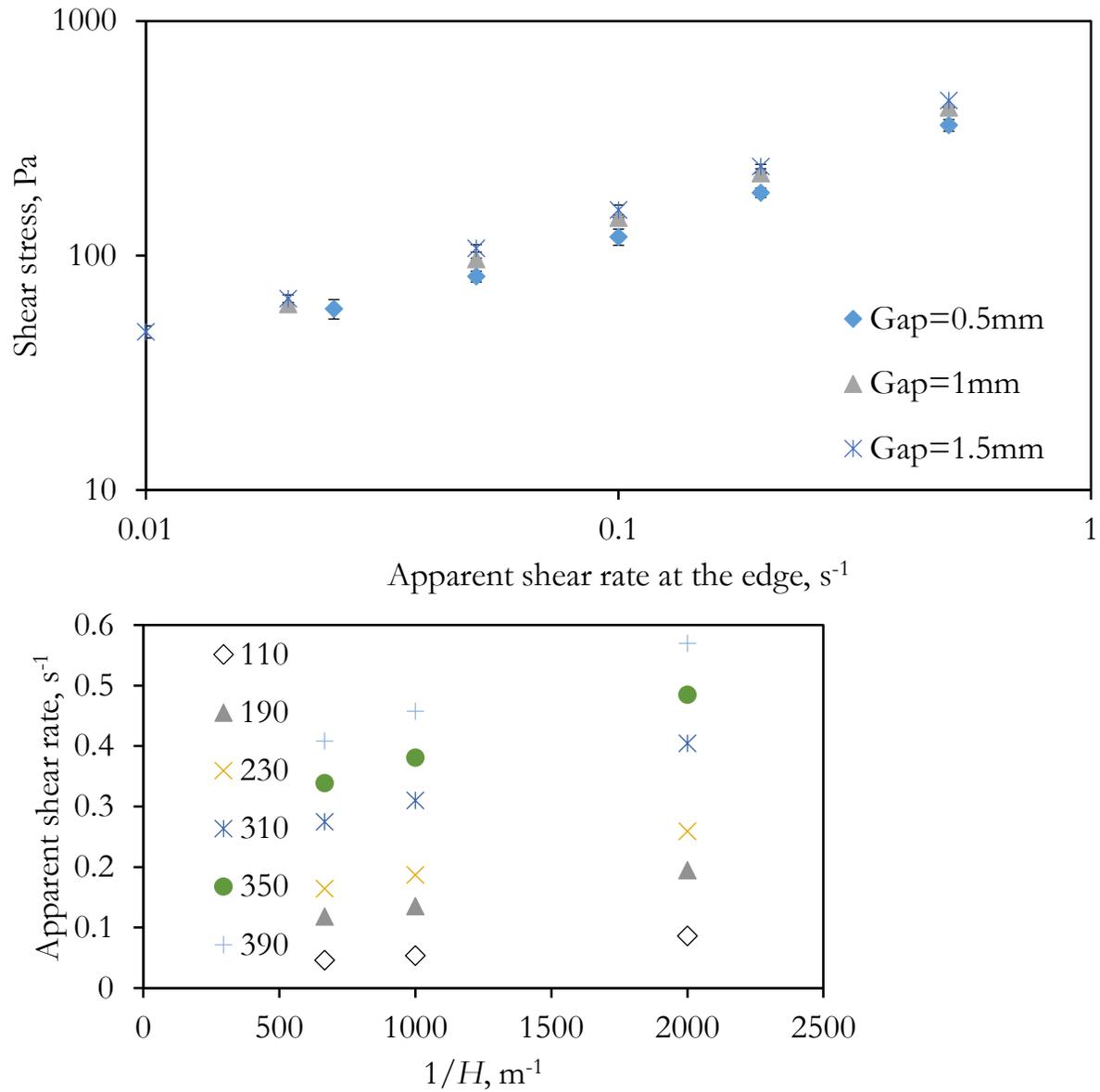

Fig. S11 Upper shows Shear stress at the edge, $\tau_R$, versus the apparent shear rate at the edge, $\dot{\gamma}_{aR}$ for different gaps or the slip corrected shear rate at the edge, $\dot{\gamma}_R$ of $\phi$=0.74 suspension in the parallel disk torsional flow. Bottom shows apparent shear rate versus reciprocal gap at indicated constant shear stress values in the parallel disk torsional flow tests.

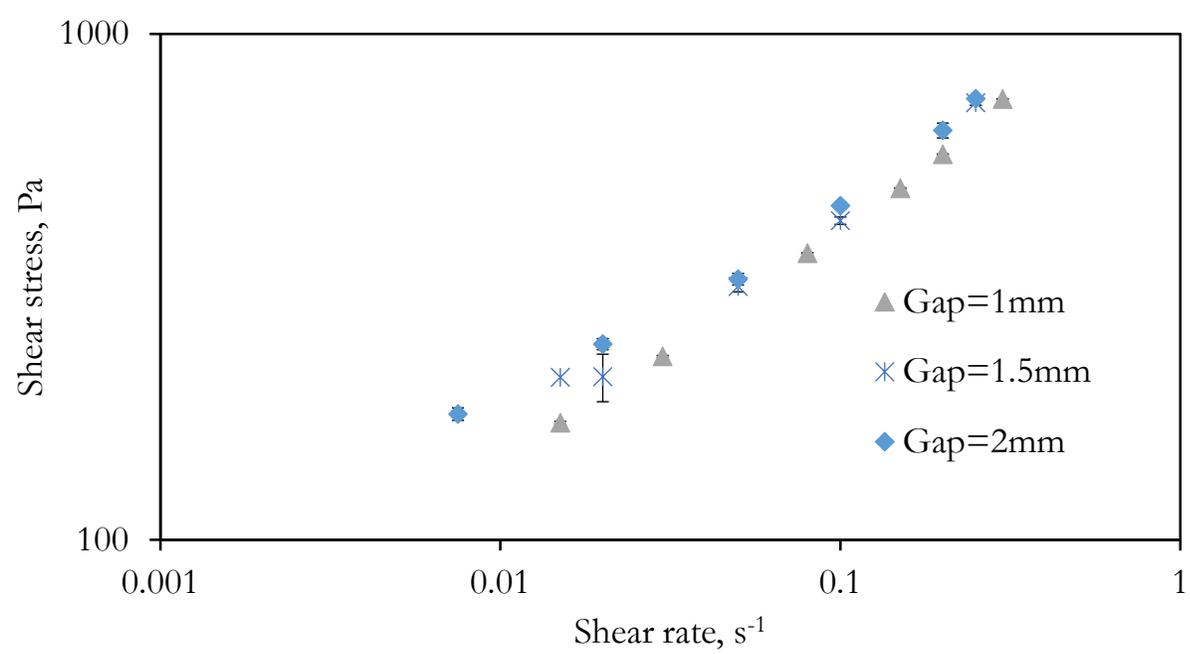

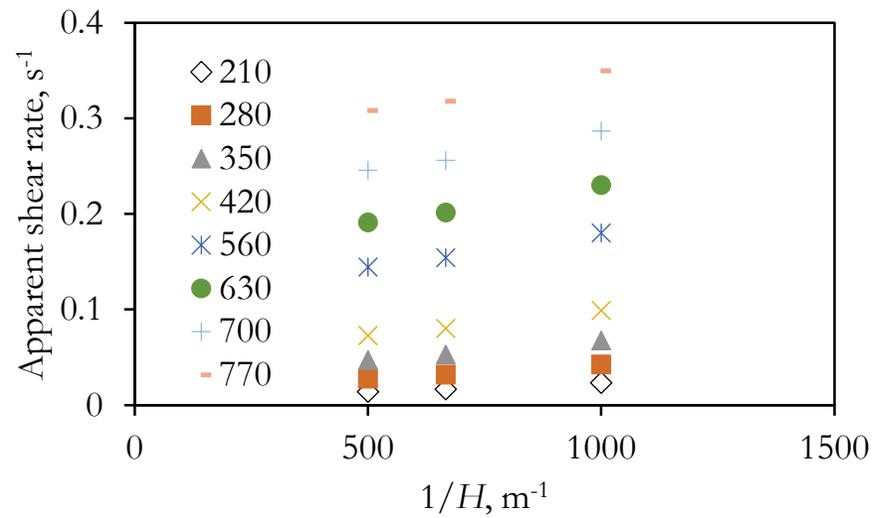

Fig. S12 Upper shows Shear stress at the edge, $\tau_R$, versus the apparent shear rate at the edge, $\dot{\gamma}_{aR}$ for different gaps or the slip corrected shear rate at the edge, $\dot{\gamma}_R$ of $\phi=0.78$ suspension in the parallel disk torsional flow. Bottom shows apparent shear rate versus reciprocal gap at indicated constant shear stress values in the parallel disk torsional flow tests.

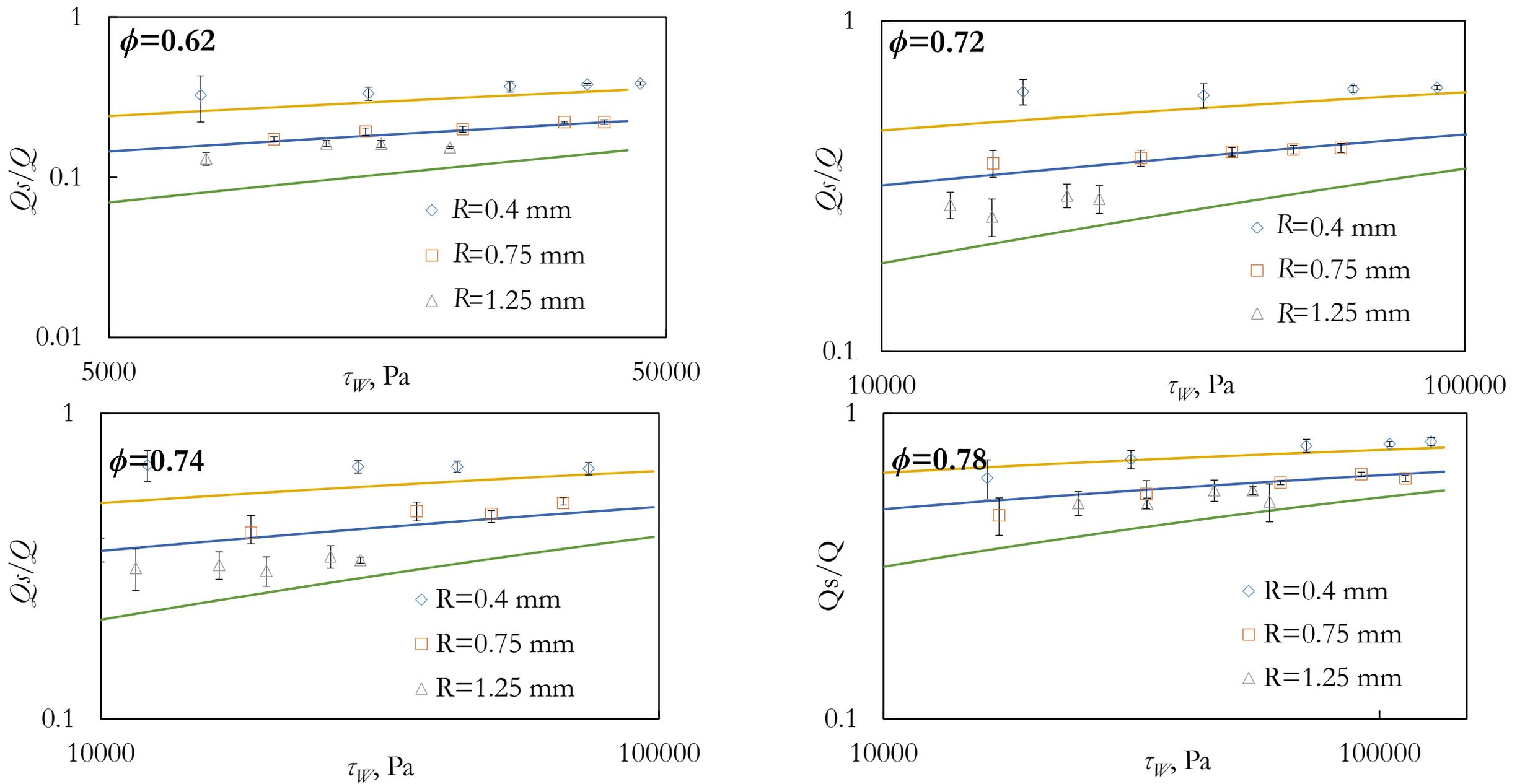

Fig. S13 Ratio of slip volumetric flow rate over total volumetric flow rate, $Q_s/Q$ versus shear stress at the wall, $\tau_W$, for capillary dies with radius of 0.4, 0.75 and 1.25 mm for $\phi$ =0.62, 0.72, 0.74, and 0.78. The lines are prediction based on Eq. (19).